\documentstyle[12pt]{article}

\newcommand{\be}{\begin{equation}}
\newcommand{\ee}{\end{equation}}
\newcommand{\bea}{\begin{eqnarray}}
\newcommand{\eea}{\end{eqnarray}}

\begin{document}
\begin{titlepage}

\begin{flushright}
{\today}
\end{flushright}
\vspace{1in}

\begin{center}

{\bf Causality, Crossing and Analyticity in Conformal Field Theories   }
\end{center}

\vspace{.2in}

\normalsize

\begin{center}
{ Jnanadeva Maharana \footnote{E-mail maharana$@$iopb.res.in}} 
\end{center}

\normalsize

\begin{center}
 {\em Institute of Physics \\
and \\
NISER\\
Bhubaneswar - 751005, India  \\    }

\end{center}

\vspace{.2in}

\baselineskip=24pt

\begin{abstract}
Analyticity and crossing properties of four point function
are investigated in conformal field theories in the frameworks
of Wightman axioms. A Hermitian scalar conformal field, satisfying
the Wightman axioms,  is considered. The crucial role of microcausality
in deriving analyticity domains is discussed and domains of 
analyticity are presented. A pair of permuted Wightman
functions are envisaged. The crossing property is derived by
appealing to the technique of analytic completion for the pair
of permuted Wightman functions. The operator product expansion of
a pair of scalar fields is studied and analyticity property of
the matrix elements of composite fields, appearing in the
operator product expansion, is investigated. An integral
representation is presented for the commutator of composite fields
where microcausality is a key ingredient. Three fundamental
theorems of axiomatic local field theories; namely,
PCT theorem, the theorem proving equivalence between PCT theorem
and weak local commutativity and the edge-of-the-wedge theorem are
invoked to derive a conformal bootstrap equation rigorously.

\end{abstract}

\vspace{.5in}

\end{titlepage}

\noindent
{\bf {1. Introduction }}.

The purpose of this article is to investigate the intimate relationships
 between causality, analyticity as well as crossing properties
of four point functions in conformal field theories. 
The seminal paper of Mack and Salam \cite{ms} led to vigorous
research activities in conformal field theory (CFT). The rapid progress 
of research 
 during that period, following work of Mack and Salam \cite{ms},
has been chronicled 
in several books and review articles 
 \cite{fgg1,mack0,fp1,fp2,todo,rev1,rev2,rev3}.\\
\noindent It is well known that wide class of  physical systems exhibit  
scale invariance at the  critical points.
The critical indices have universal character. Polyakov 
\cite{polya1}
had conjectured that such theories might be invariant under
 the conformal symmetry which encompasses
the scale invariance. 
The conformal bootstrap program was initiated by Migdal \cite{migdal},
 by Ferrara and his collaborators 
\cite{ferrara,ferrara2}, by Mack and Todorov \cite{mt}, 
and by Polyakov \cite{polya2}. The proposed conformal bootstrap
program synthesises two important ingredients: (i) conformal 
invariance and (ii) operator produce expansion.
Moreover, crossing symmetry plays a crucial 
 role in the derivation of the consistency conditions.    We shall
elaborate on some aspects of the latter and its relevance 
to our investigation in the sequel.
 The computation of the correlation functions in conformal field theories
 is of paramount
interest. 
 A lot of attention has been focused
on the formal developments and  to address important issues in
 conformal field theory itself.
The bootstrap program is understood intuitively in the following sense  
\cite{migdal,ferrara,ferrara2,mt,polya2}.
  The essential idea is  that if we consider
a 4-point correlation function  with a given ordering of operators and 
another 4-point function
where the operator orderings  are permuted then the two correlation 
functions are related
if crossing symmetry is invoked.
It becomes  transparent
when conformal partial wave expansion is implemented in each channel. 
As a result, a set of
consistency conditions emerge from the requirements of bootstrap hypothesis.
  It is accepted that the two correlation functions are analytic
continuation of each other. The interests in conformal bootstrap have been
 revived with renewed
vigor in the recent past \cite{rev1,rev2,rev3}. 
Causality, crossing and analyticity are three pillars of 
the conformal bootstrap paradigm.
\\

The  bootstrap concept stems from the S-matrix approach \cite{chew1,chew2}
 to understand  the hadronic  interactions phenomenologically in an era  
 prior to the discovery of QCD as the theory of strong interactions.
It will be useful to consider
 an illustrative example. 
 A vast  number of meson and baryon resonances were produced in high energy 
accelerators.  Let us consider meson-meson scatterings: 
(i)  $a+b\rightarrow c+d$ and (ii) $a+{\bar c}\rightarrow{\bar b}+d$.
 The first process corresponds to a direct channel 
reaction and the latter is the crossed channel one.
 In the low energy region, in each channel,  the two
 processes are dominated by exchanges of resonances.
 Notice, however, that crossing symmetry relates the two amplitudes.
 The allowed kinematical variables for each of the
 two reactions do not overlap. To state it more explicitly;
for the direct channel process, $s\ge 4m^2$ and $t\le 0$, whereas in
case of crossed channel reaction, $s\le 0$ and $t\ge 4m^2$. Here equal mass
scattering is assumed, $m$ being the mass of the external particle.
 In the fist case,
$s$ is $c.m$ energy squared and $t$ is momentum transfer squared; however,
for crossed channel reaction the roles are reversed. 
If crossing is assumed then a set of equations relating 
coupling constants and masses involved
 in two channels get related as consistent conditions.
 Thus the bootstrap hypothesis was tested phenomenologically.
 It was a challenge
 to construct a crossing symmetric amplitude and 
Veneziano \cite{gabriele} succeeded in presenting such an amplitude. 
The crossing
 symmetry was rigorously proved in the framework of axiomatic
 field theory by Bros, Epstein and Glaser\cite{beg}.
  Let us elaborate this aspect for future reference.
 The microcausality, as one of the axioms 
  of LSZ \cite{lsz} formulation,  was the main ingredients 
to demonstrate that the absorptive parts of s-channel and u-channel amplitudes
 of the 4-point functions coincided in an unphysical domain of real 
kinematical variables. It was proved rigorously from the
  theory of several complex variables for  analytic functions that 
these two absorptive amplitudes are analytic continuation of each other. The
 conformal bootstrap program was initiated to provide a field
 theoretic basis to the bootstrap paradigm of the phenomenological
S-matrix philosophy.    There  have been  stimuli
 for further research in conformal bootstrap proposal due to its applications
to variety of physical problems.
  The research in conformal field theories has spread in diverse directions
 such as in a large class of gauge theories and supersymmetric theories. 
Moreover,  studies of structure of two
 dimensional conformal field has flourished and 
 has influenced research in several directions. It is  recognized that
 conformal symmetry has played a key role in establishing 
the $AdS_5\leftrightarrow CFT_4$ correspondence
 which has led to spectacular developments in our understanding of the
 relationships between string
 theory and quantum field theories \cite{juan}: a  
 realization of the gauge-gravity duality.\\

 A very important attribute of the conformal
field theory approach is that one computes the correlations functions 
of the product of field
operators rather then the S-matrix elements. It is worthwhile to discuss 
this point.
In the axiomatic field theory approach 
\cite{lsz,wight,book1,book2,book3,book4,book5,book6},  
 it has been demonstrated that how the analyticity of the scattering 
amplitude is intertwined with microcausality. 
 The analyticity properties of  amplitude are 
rigorously proved from axioms of general field theories. 
Let us  recall some of the important results which are derived 
from a set of axioms introduced by   
   Lehmann, Symanzik and Zimmermann (LSZ)\cite{lsz}   
who laid the  foundations of general field theories.
  The S-matrix elements are computed from  the reduction technique of  LSZ.\\

  Notice, however, that
while  computing  the scattering amplitudes \cite{lsz}, 
the external particles are on the mass shell. The analyticity properties
of the amplitude are reflected through the dispersion relations. 
The dispersion relations are proved
from fundamental principles such as Lorentz invariance and microcausality.
  There are two additional  important ingredients in the LSZ
formulation besides the stipulated postulates:  (i) The existence of 
(asymptotic) {\it in} and {\it out} fields so that a complete set of
 operators can be constructed in terms of either {\it in} or {\it out} field.
 And (ii) it requires that there are massive particles in the theory. 
 Therefore, the second assumption, intuitively implies the existence of 
short range forces in the theories
 under consideration.
   The analyticity and crossing symmetry  of the S-matrix are proved
 in axiomatic frameworks for massive field theories. These results are derived in the {\it linear program} without invoking the 
 the unitarity property of S-matrix  which is a nonlinear
constraint.
 One of the most important accomplishments of the axiomatic
 approach is to study analyticity property of
the amplitude in order  to determine
the domain of holomorphy. 
 Moreover,  in order to identify the domain of analyticity 
of the amplitude in $t$, the momentum transferred squared,
 the Jost-Lehmann-Dyson 
  \cite{jl,dyson1} representation played a very crucial role. 
This paved the way
  to prove the fixed-$t$  dispersion relations. 
   However,  Bogoliubov  \cite{book6} has developed an alternative 
 technique to go off mass shell  for the external particles and 
he independently proved  fixed-$t$ dispersion relations
 for scattering  amplitudes.
 The analyticity properties
 of the S-matrix  are utilized to prove rigorous theorems, 
usually expressed as upper and lower bounds,
  on experimentally measurable parameters. Therefore, if any of those 
bounds are violated in high energy experiments
  then some of the axioms will be questioned.
 There is no evidence for violation of any of the bounds so far. \\
 
There are formidable problems
 when  the spectrum of a theory contains only massless particles.
 The pure Yang-Mills theory, Einstein's theory of gravitation
 and some conformal field theories belong to this category.   
 Our intuition guides us to the root of the problem.  Notice that,  
in such theories,  the presence of massless  particles would lead to
 long range forces. Consequently,  there are conceptual
  difficulties in
defining the asymptotic {\it in} and {\it out} states. Therefore, the 
LSZ technique is not   suitable,   while dealing  with 
some of the theories noted  above. As we shall discuss later, conformal 
field theories do not admit a discrete mass parameter. Therefore,
the LSZ formalism, as  presented in its original formulation, 
is not quite suitable in the context conformal field theories.
The  Poincar\'e group is contained in the conformal group,
 the enlarged symmetry group.
The theories are severely constrained when the conformal symmetry is enforced. 
 One notable feature is that the states are not
labelled by a $ (mass)^2$, unlike the Poincare invariant 
conventional field theories.
  We recall  that  the generators of the spacetime translations, $P_{\mu}$, 
does not commute with the Casimir operators  constructed from the generators of conformal group. Whereas, in the context of 
conventional  QFT formulations, $P^2=P_{\mu}P^{\mu}$ is a Casimir. 
The other Casimir operator is \cite{book1}  $W^{\mu}W_{\mu}$ where
$W_{\mu}=\epsilon_{\mu\nu\rho\tau}P^{\nu}M^{\rho\tau}$, 
the Pauli-Luansky vector.
The eigenvalue of the second Casimir is associated with  the helicity 
of the particle. Therefore, a state vector is represented
by its mass and helicity in quantum field theories.   
We shall discuss in the sequel, some  aspects of the unitary irreducible
 representations
of the conformal field theories which are quite different
 from those of the Poincar\'e invariant quantum field theories. \\

It is evident from the preceding discussions that computation 
of the scattering amplitudes, as stipulated by LSZ for theories
satisfying their axioms, cannot go through in a straightforward manner 
in case of CFT.  Thus it is not possible to derive an expression
for the scattering operator rigorously using the LSZ reduction
technique. Recently, however, an interesting 
development has provided evaluation of form factor and scattering
amplitude from the perspectives of LSZ technique \cite{mmp} by incorporating
a theorem from LSZ paper.
 They define form factors and scattering amplitudes in conformal
field theories. These are the coefficient of singularity of 
the Fourier transform of time-ordered correlation functions in the limit
$p_i^2\rightarrow 0$ where $p_i$ stands for four momenta of external legs.
  The form factor, $F$, is extracted from the four point
function. It is shown that $F$ is crossing symmetric, analytic and 
it admits a partial wave expansion. We note that the authors
work in Lorentzian signature metric. They obtain momentum space 
representation for the four point function from the time ordered
product of four field operators.  Indeed, this investigation is an 
endeavor to address important issues such as analyticity and crossing
symmetry from the perspective of LSZ, however, they have circumvented 
the notion of  explicitly introducing
 'in', 'out' fields as well as interacting fields in their formulation.
Moreover, the idea of interpolation, how interacting field  interpolates
into 'in' and 'out' fields, is not utilized.   \\
 
We subscribe to  the philosophy   that  it is desirable to compute correlation 
functions in conformal field theories. As a consequence, we do not
 invoke the concept of  scattering operator in our approach.
We shall unravel the relationship between analyticity, causality and crossing 
 as we proceed.
In recent times, the Euclidean formulation of conformal field theory has
been widely adopted. There are certain advantages in this approach while
the position space description is utilized. The merits of Euclidean
space formulation and its power has been discussed in \cite{rev1,rev2,rev3}.
On the other hand the Minkowski space (Lorentzian signature metric) formulation
has its own merits as has been discussed in \cite{bg,mg}. We feel that
the choice of the Lorentzian metric is more suitable for 
establishing relationship between causality and analyticity as is evident
from the rigorous results derived in axiomatic field theories. As a motivation,
we present an illustrative example which is a very simplified version
of Toll's work in the context of dispersion relation \cite{toll}.\\

Let us consider a wave packet, $\psi(z,t)$, moving with velocity $c$ along 
the $z$-axis. 
The target is located at the point $z=0$. 
The Fourier transform of $\psi(z,t)$ is $f(\omega)$ and we can write 
\bea
\label{sec1.1}
\psi(z,t)={1\over{{\sqrt 2\pi}}}\int_{\infty}^{+\infty} d\omega 
f(\omega)e^{i\omega({{z\over c}}-t)}
\eea
The spherically symmetric scattered wave has the form
\bea
\label{sec1.2}
S(r,t)= {{1\over{r{\sqrt{2\pi}}}}}\int_{-\infty}^{+\infty}{\tilde S}(\omega)
f(\omega)e^{i\omega({{z\over c}}-t)}
\eea
Here $r$ is the radial distance from $z=0$.  We may invert eq. (\ref{sec1.1})
to get
\bea
\label{sec1.3}
f(\omega)={{1}\over{\sqrt{2\pi}}}\int_{-\infty}^{+\infty}dt\psi(0,t)
e^{i\omega t)} 
\eea
We recognize that if the wave packet does not reach the origin, $z=0$, then
there will be no observation of scattered wave; in other words if
\bea
\label{sec1.4}
\psi(0,t)=0,~~{\rm for}~~t<0
\eea
then $f(\omega)$ is regular for ${\rm Im}~\omega>0$; i.e. it is regular
on the upper half of the complex $\omega$ plane. \\
For us, causality, in the present context, means no scattered wave would
arrive at a distance $r$ from the origin after an interval ${r\over c}$
when the incident wave has not reached $z=0$. 
\bea
\label{sec2.5}
S(r,t)=0,~~{\rm for}~~(ct-r)<0.
\eea
Let us take inverse Fourier transform of (\ref{sec1.2}). We conclude that
$f(\omega){\tilde S}(\omega)$ is analytic in the upper half $\omega$ plane;
i.e. it is analytic when ${\rm Im}~\omega>0$. Therefore, ${\tilde S}(\omega)$,
the Fourier transform of the amplitude, is analytic for  ${\rm Im}~\omega>0$.
This property is exploited to write a dispersion relation. In the
context of special relativity velocity of light, $c$, is the limiting
velocity and therefore, two points separated by spacelike distance are 
causally disconnected. In the context of quantum theory, if there are two
Hermitian operators located at spacelike distance we can make observations
simultaneously. \\

In relativistic quantum field theories the causality is stated to be
$[O(x),O(x')]=0,~{\rm if}~(x-x')^2<0$ where $O(x)$ and $O(x')$ are two local
operators. The analyticity properties of scattering amplitude crucially
depend on the axiom of microcausality. The dispersion relations rest
primarily on this axiom.  
In order to
understand concepts such as causality and crossing intuitively
 it is desirable
 to work in a Minkowski spacetime manifold with
a Lorentzian metric. Our  adopted signature for the flat space metric is
$g_{\mu\nu}= {\rm diag}~(+,-,-,-)$.\\

  Moreover, we shall attempt to reveal how these three fundamental
 properties are intertwined in the Wightman formulation of field theories
in the context of conformal field theories.
 The necessary ingredients,  to explore these aspects would be
 discussed at an appropriate juncture.
The correlation functions   are the vacuum expectation values of field or 
local operators i.e. the Wightman functions. However, it must be borne 
in mind that not all
conformal fields fulfill the requirements of Wightman axioms.
 We shall discuss this aspect in the later part of this section.  Another advantage of adopting the 
Wightman's formulation is that the Wightman functions are boundary
values of analytic functions. It facilitates to establish 
relationship between causality and analyticity and subsequently paves the
way to an understanding of crossing once we define analytic functions
from the Wightman functions. 
\\ 

The operator product expansion (OPE) has played a very important role 
in the advancements of
conformal field theories in recent years.
 Wilson \cite{wilson} 
pioneered the  technique of operator product expansion in QFT. His
 seminal works  have  profoundly influenced research in diverse branches of 
physics. 
 Wilson's theory was further advanced  rigorously  
by Wilson and Zimmermann \cite{wz} 
 in  the  frameworks  of Wightman axioms \cite{book2}. It was shown
 that under
certain conditions, the operator product expansion can be derived  
giving complete information on short distance behaviors.
 They provided procedures to 
 construct composite field operators from the perspectives of 
Wightman axioms,
 supplemented by extra hypothesis, relevant for operator product 
expansions in QFT.    It was argued,  by Wilson and Zimmermann   
  that , when the  products of local operators are envisaged  
singularities invariably
occur as separation between the two operators tends to vanish.
 These singularities appear as a consequence of relativistic
 invariance and positive definite metric in Hilbert space \cite{lehmann}.  
Subsequently, Otterson and Zimmermann \cite{oz}, 
 concentrated on Wilson expansion for product of two scalar  field operators: 
$A(x_1)A(x_2)$. They demonstrated, following the approach of \cite{wz},
  that this 
operator product could be used to define composite local operators
 i.e. local with respect to $(x_1+x_2)/ 2$. The OPE also  depends,
in addition, on a vector, $\zeta^{\mu}$, which is proportional to 
distance $(x_1-x_2)^{\mu}$ between $A(x_1)$
and $A(x_2)$. They investigated locality and
analyticity properties of appropriate matrix elements  from the Wightman 
axiom point of view.
 Their conclusions are based on rigorous techniques and   are   quite
robust.  \\

We focus our attentions to study 
the consequences of locality and analyticity in conformal field theories and 
it will be carried out  in the light of
the works of \cite{wz} and \cite{oz}. Moreover, the 
procedures of \cite{oz} is appropriately adopted for the conformal
field theories.  
In the context of CFT, the OPE imposes severe constraints on 
the structural frameworks of the  theories. Furthermore, it enables us
 to extract important conclusions without resorting to any specific model 
i.e. without introducing
 a Lagrangian density or an action. We are aware that the issue
of convergence of OPE in conformal field theory is quite
pertinent. The derivation of the conformal bootstrap equation
is intimately connected with the convergence of conformal partial
wave expansion. It is worth while to note that, in the context of
conformal field theories, supplementary postulates are needed
in addition to the Wightman axioms. 
 Microcausality is a cardinal principle of axiomatic 
 local field theories. Therefore, for a theory with three important ingredients such as
  (i) an enlarged symmetry like  the conformal invariance, (ii)
 microcausality and (iii) the  power
 of operator product expansion; some of the important and most 
general attributes are extracted. 
 For example, the two point function and the three point function
 of a CFT  are determined  up to  multiplicative constant factors.
 Furthermore, 
very important characteristics  and structure of four point functions 
are understood 
 from these ingredients.\\  
 
The crossing symmetry, in the context of conformal field theories in 
conjunction with the power of operator product expansion,
 is generally discussed in the  coordinate space description. 
Let us consider the three point function for a scalar field  where the field
 operators are located at spacetime points $x_1$, $x_2$ and $x_3$  and 
is defined as follows.
 $W_3(x_1,x_2,x_3)=<0|\phi(x_1)\phi(x_2)\phi(x_3)|0>$. 
If we consider a permuted field configuration then
  $W_3(x_1,x_3,x_2)= <0|\phi(x_1)\phi(x_3)\phi(x_2)|0>$; 
  the two correlation functions are equal i.e. 
$W_3(x_1,x_2,x_3)=W_3(x_1,x_3,x_2)$
 when $(x_2-x_3)^2<0$; that is when $x_2^{\mu}$  and $x_3^{\mu}$ 
separation is spacelike.   The essential point to note is that
  it is necessary to prove that $W_3(x_1,x_2,x_3)$ and   $W_3(x_1,x_3,x_2)$
  are analytic continuation of each other. Similar arguments can
  be invoked for the four point functions as well
 as for $n$-point functions while discussing crossing. \\
  
We hold  the view that the study of the analyticity properties of
the correlation functions in CFT is best accomplished in 
the  Wightman's formulation. The Wightman  functions are considered to be
  boundary values of analytic functions of several 
complex variables. Let us consider an n-point Wightman function
  $W_n(x_1,x_2,....x_n)=<0|\phi(x_1)\phi(x_2).......\phi(x_n)|0>$. 
It is the boundary value of an analytic function. If we permute location of
  field operators then another corresponding  Wightman function would 
be defined which would be boundary value of another analytic
  function.  Our task, in the context of CFT,
 is to investigate analyticity properties of 
four point Wightman functions.   
  Consider the four point function 
$W_4=<0|\phi(x_1)\phi(x_2)\phi(x_3)\phi(x_4)|0>$. It  is crossing 
symmetric if we compute $W_4$ using
 standard OPE technique as long as all four points are separated 
by spacelike distances, modulo the issues related to the
convergence of the operator product expansions.
 Indeed, the consistency requirements for the four point
 amplitude, computed by contraction of pair of field (for different 
configurations), leads to the so called {\it conformal bootstrap program}
(see section 3 for the work in this direction). 
 We would like to address the following question: how far we can proceed within an axiomatic framework to infer about crossing and analyticity properties
 of the correlation functions of conformal field theories?\\

 Our goal is modest. We intend to study only a class of four 
dimensional conformal field theories which  respect Wightman axioms.
  The generators of the conformal group are: (i) the ten generators of 
the Poincar\'e
group i.e. spacetime translations and Lorentz transformations 
respectively denoted by $P_{\mu}$ and $M_{\mu\nu}$. (ii)  
The five additional
generators are: the dilatation operator, D,  ( associated with the scale 
transformation) and the four generators corresponding to special conformal
transformations are denoted by $K_{\mu}$. Thus there are 
altogether fifteen generators in $D=4$. The scalar field, $\phi(x)$, 
transforms as
\bea
\label{conf1}
[P_{\mu},\phi(x)]=i\partial_{\mu}\phi(x),~~
[M_{\mu\nu},\phi(x)]=i(x_{\mu}\partial_{\nu}-x_{\nu}\partial_{\mu})\phi(x)
\eea
\bea
\label{conf2}
[D,\phi(x)]=(d+x^{\nu}\partial_{\nu})\phi(x),~~[K_{\mu},\phi(x)]=i(x^2\partial_{\mu}-2x_{\mu}x^{\nu}x_{\nu}-x_{\mu}d)\phi(x)
\eea
where $d$ is the scale dimension of the field $\phi(x)$.
  The above transformation rules are modified appropriately when
 we consider  tensor
fields. The fifteen generators of the conformal group satisfy the algebra
 of $SO(4,2)$. There are
three Casimir operators which are constructed from  these generators. 
We give below the expressions for the Casimir operators for the sake of
completeness.
 \bea
 \label{conf3}
 C_2={1\over 2}M_{\mu\nu}M^{\mu\nu}-K_{\mu}P^{\mu}-4iD-D^2
 \eea
 \bea
 \label{conf4}
 C_3= -{1\over 4}\bigg(W_{\mu}K^{\mu}+K_{\mu}W^{\mu}\bigg)-{1\over 8}\epsilon_{\mu\nu\rho\tau}M^{\mu\nu}M^{\rho\tau}
 \eea
 \bea
 \label{conf5}
C_4= &&{1\over 4}\bigg[K_{\mu}K^{\mu}P_{\nu}P^{\nu}-4K_{\mu}M^{\mu\nu}M_{\nu\rho}P^{\rho}-4
K_{\mu}M^{\mu\nu}P_{\nu}(D+6i)\nonumber\\&&
+{3\over 4}(M_{\mu\nu}M^{\mu\nu})^2+{1\over{16}}(\epsilon_{\mu\nu\rho\tau}M^{\mu\nu}M^{\rho\tau})^2\nonumber\\&&
+M_{\mu\nu}M^{\mu\nu}(D^2+8iD-C_2-22)-D^4-16iD^3\nonumber\\&&
+80D^2+128iD+36C_2-16iC_2D-2C_2D^2 \bigg]
\eea
where $W_{\mu}={1\over 2}\epsilon_{\mu\nu\rho\tau}P^{\nu}M^{\rho\tau}$. 
The expressions for the Casimir operators  for 
four dimensional Minkowski space, as   given above, 
 are according to  the notations and conventions of 
Fradkin and Plachek   \cite{fp2,fp3}.\\ 

Let us discuss Wightman approach in the context of conformal field theories.  
The analyticity properties structure of conformal field theories
was systematically investigated by L\"uscher and Mack \cite{luscher} 
 extensively. We shall discuss some
of the salient aspects in the next section. Let ${\tilde\phi}(p)$ denote the
Fourier transform of $\phi(x)$. Notice that  the Fourier transform, 
${\tilde\phi}(p)$, of  every conformal field does not satisfy 
spectrality condition of
Wightman axioms \cite{wight}; i.e. $p^2\ge 0$ and $p^0\ge 0$. A special type 
of conformal fields fulfill the spectrality condition. These are designated
as nonderivative fields \cite{mack1}. They satisfy the constraint 
$[\phi(0), K_{\mu}]=0$. Note that $x$-dependence of the field can be 
accomplished through the transformation: $\phi(x)=e^{-iP.x}\phi(0)e^{iP.x}$. We shall consider the nonderivative conformal fields, 
denoted by $\phi(x)$,  throughout this investigation. 
 A state belonging to a unitary  irreducible representation is specified 
by appropriate quantum numbers of the covering 
 group of $SO(2,4)$.
  The classification of the representations have been studied several 
decades ago \cite{yao1,yao2,yao3,mack2}. 
A desirable and useful line to pursue is to envisage the 
universal covering group of $SO(4,2)$ to be $SU(2,2)$. We mention below
the following postulates of the Wightman
 formulation which are quite important in the context of conformal
field theories. A summary of Wightman axioms and its implications
will be presented in the next section. The two important ingredients are:
  \\
{\it (i)The vacuum}: The vacuum is denoted as $|0>$. The vacuum is 
conformally invariant and therefore,   the generators of the conformal group
annihilate the vacuum.  
\\
{\it ( ii) The Hilbert Space}:  
A radically new approach is invoked in the study of conformal field 
theories (CFT).  The structure of the Hilbert space,
in the case of conformal field theories, is different which would be alluded to
later \cite{fp2}. 
\\

We discuss briefly the correspondence between local operators and states. 
This concept plays a crucial role in understanding
of the structure of the Hilbert space in conformal field theories. The state may be constructed either in the
momentum space representation or in the coordinate representation as well. Consider a local operator, ${\cal O}(x)$, acting on the vacuum.
The operator ${\leftrightarrow}$ state correspondence is
\bea
\label{conf2a}
{\cal O}(x)|0>=|{\cal O}(x)>
\eea
In fact we could consider the operator at $x=0$, i.e. ${\cal O}(0)$ and generate $x$-dependence by a translation
operation.  The momentum space state vector is
\bea
\label{conf2b}
|{\cal O}(p)>=\int d^4xe^{ip.x}|{\cal O}(x)>
\eea
Therefore, the vacuum expectation values of product of 
operators can be evaluated either for those in $p$-space
representations or those in $x$-space representations.
  In the framework of conformal field theories, we have a set of
fields. 
One does not
  envisage  a model or a Lagrangian density with a single field or
 a set of fields while investigating  general structure of CFT.
 Thus the Hilbert space is constructed from such a set of fields:
\bea
\label{conf3}
\{{\Phi_m}\}:~{\Phi_0(x)},{\Phi_1(x)},.......
\eea
Each field, ${\Phi_m(x)}$, carries a scale dimension, $d_m$,
 and might be endowed with its own tensor structure; moreover,
it  might be characterized with   internal
quantum numbers.
 The product of two conformal fields is expanded in terms of 
complete set of local fields with C-number coefficients. Therefore, we need
 an infinite set of local fields which belong to unitary irreducible
 representation of the conformal group. Generically we express the product 
 of a pair as 
\bea
\label{conf4}
{\Phi(x_1)}{\Phi(x_2)}=\sum_{m=0}^{\infty}A_m\int C_m(x:x_1,x_2){\Phi_m}(x) d^4x
\eea
where $\{ \Phi_m(x) \}$ are the set of fields belonging to irreducible 
representation of the conformal group and $C_m(x;x_1,x_2)$,
the  C-number functions, which
have singular behavior in  the short  distance limit. $A_m$ are a set of 
constants, may be interpreted as coupling constants and these are determined 
from the
dynamical inputs of the specific theory under considerations. 
We may associate a state vector with each of the fields in 
$\{\Phi_m(x) \}$ from the hypothesis of state $\leftrightarrow$ operator 
correspondence alluded to already. A state vector is  created
when a field operator acts on  
 the conformal vacuum as noted earlier. Thus a state in the Hilbert 
space is defined   
\bea
\label{conf5}
|{\Phi}_m>=\Phi_m|0> 
\eea
The full Hilbert space is decomposed into direct sum of mutually 
orthogonal spaces since two normalized states $|\Phi_m>$ and $|\Phi_n>$ 
with respective scale
dimensions $d_m$ and $d_n$ are orthogonal i.e.
 $<{\Phi}_m|{\Phi}_n>=\delta_{m,n}$.  In fact
 each of the states belonging
to unitary irreducible representations of the conformal group 
constitute subvector spaces. Therefore, the full Hilbert space, 
${\bf{\cal H}}$,
decomposes into direct sums as
\bea
\label{conf5a}
{\bf{\cal H} }=\oplus {\cal H}^{\chi}
\eea
 ${\cal H}^{\chi}$ is the subspace where the complete set 
of state vectors are created by complete set of fields ${\Phi}^{\chi}$ which
belong to an irreducible representation of the conformal group.
 Here $\chi$ stands, collectively, for all the quantum numbers that
 characterize
the irreducible representation. 
   A noteworthy feature is the closure of the algebra in the sense that 
the product of two fields can expanded
in terms fields of conformal field theory: 
symbolically;  $\Phi_m\Phi_n\sim \sum_l A_{mn}^l\Phi_l$; where $A_{mn}^l$
are set of C-number coefficients. \\

The choice of Lorentzian signature to study the properties of conformal
 field theories has certain advantages as has been noted in the early
phase of the research in CFT \cite{ms,fgg1,migdal,polya2,fp2}. 
 Moreover, the Wightman function as VEV of the product
of field operators and their variances in the form of the VEV of R-products 
or A-product of fields have been utilized for good reasons.  For example,
Polyakov \cite{polya2}  had emphasized  for adopting R-product and 
A-product of a pair of fields to evaluate corresponding Wightman function.
Its importance was noted while computing the discontinuity of four point function in connection with the conformal bootstrap equations.  
\\

Let us consider the three point Wightman function 
 for  scalar fields with scale dimension $d$. Our interest lies
in the study of analyticity properties. Therefore, we adopt 
the $i\epsilon$ prescription for all the Wightman functions
in what follows: ( see more details in subsequent  next section when we
complexify the coordinates).
After  suitably implementing the conformal transformations  
on each of the scalar fields,  the three point Wightman function 
assumes the following form.
\bea
\label{conf7}
W_{123}(x_1,x_2,x_3)=&&<0|\phi(x_1)\phi(x_2)\phi(x_3)|0>\nonumber\\&&
=g_3(x_{12}^2-i\epsilon x_{12}^0)^{-d/2}(x_{23}^2-i\epsilon x_{23}^0)^{-d/2}(x_{13}^2-i\epsilon x_{13}^0)^{-d/2}
\eea
The above expression is presented for the scalar field, $\phi(x)$; however, 
the expression for three point
function of three arbitrary field is already known in the literature. 
Here    $g_3$ is interpreted a the coupling constant.
 \\
 
When fields  are separated by 
spacelike distances i.e. $(x_1-x_2)^2<0$, $(x_2-x_3)^2<0$ and $(x_1-x_3)^2<0$
then the   Wightman functions are related to each other: 
$W_{123}(x_1,x_2,x_3)=W_{132}(x_1,x_3,x_2)=W_{312}(x_3,x_1,x_2)$
 as noted earlier; expressed
  explicitly
\bea
\label{conf8}
<0|\phi(x_1)\phi(x_2)\phi(x_3)|0>=<0|\phi(x_1)\phi(x_3)\phi(x_2)|0>=<0|\phi(x_3)\phi(x_1)\phi(x_2)|0>
\eea
This  is a consequence of the axiom  of microcausality. Indeed, 
 the equation relating different
3-point function (\ref{conf8}) for the permutations of the field 
operators is the crossing relation for vertex functions. In order to recognize
the importance of the study of analyticity, let us examine the expressions 
for a pair of three point functions.
In the case of the Wightman functions these are boundary values of analytic
 functions (see later).  
Consider the   expressions for  $W_{123}$ and $W_{132}$ 
\bea
\label{conf10}
W_{123}=g(x_{12}^2-i\epsilon x_{12}^0)^{-d/2}
(x_{23}^2-i\epsilon x_{23}^0)^{-d/2}(x_{13}^2-i\epsilon x_{13}^0)^{-d/2}
\eea
and
\bea
\label{conf11}
W_{132}=g(x_{13}^2-i\epsilon x_{13}^0)^{-d/2}
(x_{12}^2-i\epsilon x_{12}^0)^{-d/2}(x_{32}^2-i\epsilon x_{32}^0)^{-d/2}
\eea 
  We note that $W_{123}$ and $W_{132}$ do not necessarily 
coincide when $x_{12}$,
$x_{23}$  and $x_{13}$ are  timelike. Let us 
  compare each term of (\ref{conf10}) and (\ref{conf11}) on
 the $r.h.s$ and especially focus attention on the second
term of (\ref{conf10}) and the third term of (\ref{conf11}).
 The former is $(x_{23}^2-i\epsilon (x_2-x_3)^0)^{-d/2}$ whereas the
 latter is $(x_{32}^2 +i\epsilon(x_{23})^0)^{-d/2}$.
 Therefore, when we approach the real axis in the complex plane we must approach it from opposite directions. 
 Thus in order to provide a proof of crossing
we have to appeal to the the method of analytic completion. In recent years,
the study of the analyticity properties in CFT have attracted attentions
\cite{kz,hkt,hjk,chp,sch,bg,mg}.  \\

It is more interesting to consider the 
a pair of 4-point Wightman function, with  the $i\epsilon$ prescription, 
which have the following form  in the convention of \cite{rev3}
 \bea
 \label{conf7a}
 W_4(x_1,x_2,x_3,x_4)=&&<0|\phi(x_1)\phi(x_2)\phi(x_3)\phi(x_4)|0>\nonumber\\&&=[{1\over{({\tilde{z}}_{12}^2-i\epsilon{\tilde z}_{12}^0)
 ({\tilde{z }_{34}^2-i\epsilon}{\tilde z}_{34}^0})}]^{d}{\cal F}(Z_1,Z_2)
 \eea
for the case when we have only one scalar field, $\phi(x)$, with scale
dimension $d$; and 
 ${\tilde z}_{12}=x_1-x_2, {\tilde z}_{34}=x_3-x_4, {\bar z}_{13}=x_1-x_3, {\tilde z}_{24}=x_2-x_4, {\tilde z}_{14}=x_1-x_4, {\tilde z}_{23}=x_2-x_3$. 
 Here  $Z_1$ and $Z_2$ are the cross ratios
 \bea
 \label{7b}
 Z_1={{({\tilde z}_{12}^2-i\epsilon{\tilde z}_{12}^0)({\bar z}_{34}^2-i\epsilon{\tilde z}_{34}^0)}\over{({\tilde z}_{13}^2-i\epsilon{\tilde z}_{13}^0)
 ({\tilde z}_{24}^2-i\epsilon {\tilde z}_{24}^0)}}
 \eea
 and 
 \bea
 \label{7c}
 Z_2={{({\tilde z}_{14}^2-i\epsilon{\tilde z}_{14}^0)
({\tilde z}_{23}^2-i\epsilon{\tilde z}_{23}^0)}
\over{({\tilde z}_{13}^2-i\epsilon{\tilde z}_{13}^0)
 ({\tilde z}_{24}^2-i\epsilon {\tilde z}_{24}^0)}}
 \eea
  $\cal F$ is a function which depends on cross ratios  and its 
form is determined by the model under considerations. 
   The discussions 
 of crossing operation is  to be treated with care. For example, 
in order to establish crossing, it is desirable to show how 
$W_4(x_1,x_2,x_3,x_4)$ is analytically
 continued to $W_4(x_1,x_2,x_4,x_3)$ . They coincide  for real $x_2$ and 
$x_3$ when  $(x_3-x_4)^2<0$ i.e. when the two spacetime points are separated 
by spacelike
 distance. 
The holomorphic properties of scattering amplitudes, in the momentum space
representation, has been thoroughly studied in the past. We shall
focus primarily on the momentum space representation of
three point functions in Sction 2 and study their analyticity
properties.
 \\

The article is organized as follows. In the next section, Section 2, 
we first  present a summary of the important results in
 the Wightman formulation of general field theory.
   Next, in this section,  we study the domain of holomorphy of four 
point functions and 
 a simple method is adopted to accomplish   analytic completion. 
The domain of analyticity of three point Wightman function was 
investigated in detail by
  K\"all\'en and Wightman \cite{kw}.
  The problem of analytic completion for four point function deserves some
discussions. There are several attempts to accomplish analytic 
completion for four point function \cite{kt,kallen,russian} and there 
has been some  progress. We mention,
in passing, that these endeavors have not been able to accomplish
as much as what have been  achieved exhaustively  for 
the three point function by K\"all\'en and Wightman  \cite{kw}.
Our goal, in this article, is to address issues associated with crossing.
Therefore, following the line of arguments and technique of
 our previous work \cite{mpla}, in the context of three point function,
we investigate crossing for the four point function. We employed the
analytic completion technique for 
a pair of permuted Wightman four point functions
and showed that they are analytic continuation of each other. Thus all 
Wightman functions are shown to be 
 analytic continuation of each other  pairwise.
However, the holomorphic envelope might be much larger. 
 We shall next  consider the four point function for a nonderivative 
  conformal field   theory.  It will be continuation of our previous
 effort to study crossing symmetry.
   We  investigated the analyticity property and crossing for three point
 Wightman function in conformal 
   field theories \cite{mpla}. We applied the technique to a pair
 of Wightman functions taken at a time. 
   In other words if we permute a pair of field to obtain a new three point
 function from a given configuration 
   it possible to implement analytic continuation in a simple manner.
 Our work synthesized work of Dyson \cite{dyson111} and its
   modified version adopted by Streater \cite{streater}.
Dyson \cite{dyson111} had obtained a representation for double commutator of 
three scalar fields: $A(x),B(x)$ and $C(x)$. This proposal
was suitably modified by Streater \cite{streater}  
to implement the technique of analytic
completion for a pair of Wightman functions. We adopted Streater's 
prescription to prove crossing for a pair of 
three point functions by in 
 CFT applying method of analytic completion. We recall, 
if we permute a pair of field to obtain a new three point
function from a given configuration then, it was shown that,
 the two vertex functions are 
analytic continuation of each other. One remark is quite pertinent at 
this stage. The Wightman functions are boundary values of analytic functions
of several complex variables of complexified coordinates as will be defined
in the next section. The important point is these analytic functions are
defined over a domain known as extended tubes (see Section 2 for details).
There is a Wightman function defined for a given ordering of fields. There
a Wightman function for each permutation of the field ordering. Moreover,
corresponding to each Wightman function there is an analytic function,
defined over an extended tube and
the Wightman function is its boundary value. The domain of holomorphy is the
union of the extended tubes on which the permuted Wightman functions are 
defined. It is argued that the set of analytic functions so obtained are
analytic continuation of each other. Notice that  Wightman function,
defined as vacuum expectation value of product of field operators is
not related to another Wightman function obtained from permutation of the 
fields.
  Thus  it might not be possible, in our approach, to identify the envelope of
holomorphy when the union of domain of analyticities of 
all permuted Wightman functions taken. The reason
is that the envelope of holomorphy might be larger than the 
union of the domains of analyticities.  We shall elaborate
on this aspect in Section 2. We
carry out a detailed study, adopting an argument of Jost which provides
a relationship between the retarded three point function and the three point
Wightman function. This result is appropriately tailored for conformal
field theory. 
\\

Section 3 deals with operator product expansion in
 conformal field theories adopting the formalism of Wilson and Zimmermann.
The prescription laid down by Otterson and Zimmerman \cite{oz} 
for operator product expansion of two field
operator  in the frameworks of Wilson is quite rigorous. 
We employ the procedure for operator product
expansion of  nonderivative conformal fields. The matrix elements 
of the composite field are investigated
to examine consequences of microcausality.  It is shown how a 
theorem, analogous to the Jost-Lehmann-Dyson
theorem can be proved. Furthermore, the analyticity property of 
the matrix elements are derived to prove
crossing for the four point function. 
 In the later part of third section (in the subsection), 
we appeal to PCT theorem to derive conformal
bootstrap equation in a novel way. The PCT operation, applied to a four point
 Wightman function,  transforms into another Wightman function.
 These two four point functions are equal if PCT is a symmetry.  The
 equivalence between PCT theorem and weak local commutativity (WLC)
plays a very important role in the derivation of the equation. First, the
 conformal partial wave (CPW) expansion technique is  employed
to the each of the four point equations which are related
 by WLC. The two functions coincide at the Jost point for real values
of spacetime coordinates when their separation is spacelike.
 As we shall explain in Section 3, the
 two Wightman functions are boundary values of analytic functions
 and they are holomorphic as complex valued functions. We argue that
 these two four point functions are analytic continuation of each other.
Thus, we present a rigorous derivation of the
 conformal bootstrap equation. In a short communication \cite{jmplb},
we have reported part of this result.
 The summary of our work and conclusions are incorporated in Section 4.

\bigskip
\noindent
{\bf{2.   The  Wightman Functions in Conformal Field Theory}}.

\bigskip
\noindent We study the crossing and analyticity properties of the four point
function of a conformal field, $\phi(x)$,  in this section. The crossing
symmetry relates two permuted Wightman functions. We have noted, in Section 1,
 the motivations for the study of crossing in CFT. It was pointed out that
two Wightman functions, where a pair of fields get interchanged,
 coincide  when the corresponding pair of spacetime coordinates are
separated by spacelike distance. The conformal bootstrap equation
is derived under such a condition.
 This is not the full story. The proof of crossing in QFT, from
the axiomatic stand point, demands more. It is necessary to prove that
the  pair of functions have analytic continuations to their domain of
holomorphy although the pair coincide at certain spacetime point.
Therefore, the domain of holomorphy of each of the functions have  to be
identified. This is the first task. The role of OPE in CFT has been emphasized
already. 
 In turn, as was noted in the previous section, the Hilbert space
structure of CFT is different 
from that of  the conventional (axiomatic) fields
theories. We discuss this aspect very briefly in the sequel.
 We have emphasized the
relationship between microcausality and analyticity. A simple way
to bring out their intimate relationship is to consider crossing
property of the three point Wightman function. This problem has been
studied recently \cite{mpla}. We briefly recall essence of this 
 work, later in this section and incorporate our new results. 
A momentum space formulation is presented and analyticity properties are
studied by appealing to works of Jost and Ruelle (see subsection 2.2
for details).
It
will set a background to study relation between microcausality and analyticity
properties of the four point function. Moreover, we invoke the arguments
of \cite{mpla} to prove crossing for four point function.   \\

\bigskip

\noindent{\bf 2.1 Properties of Wightman Functions}.

\noindent
The Wightman functions \cite{wight} 
are vacuum expectation values of product of field operators.
They are envisaged as boundary values of analytic function of several
complex variables.
Wightman \cite{wight2} 
has argued that if all the vacuum expectation values
$<0|\phi(x_1)\phi(x_2)...\phi(x_n)|0>$ for all the permutations of
$\phi(x_1)....\phi(x_n)$
are given then the operator $\phi(x)$ is determined.  Our intent is to consider the Wightman functions for the
scalar conformal field $\phi(x)$
and study their analyticity properties and crossing symmetry. The  axioms are:
\\
{\bf A1}.  Invariance of the theory under proper inhomogeneous Lorentz group.\\
{\bf A2}. The existence of vacuum. There exists a Hilbert space,
${\bf {\cal H}}$,  spanned by the physical state vectors. The states have
nonnegative energy spectrum i.e. $p_0\ge 0$ and $p^2\ge 0$. There exists a
vacuum, $|0>$, the lowest energy state
such that if $P_{\mu}$ is
the energy momentum operator, $P_{\mu}|0>=0$. The vacuum is unique and stable.
The vacuum is annihilated by all the generators of the conformal
group. \\
{\bf A3}. There exist field operators which are  tempered. In other words,
vacuum expectation values of operators are tempered
distributions in the Schwartzian sense.\\
{\bf A4}. Local commutativity. Expressed in another way; local operators
commute (for bosons) or anticommute (for fermions) when they are separated
by spacelike distance. For bosonic local operators,
\bea
\label{sec2.1}
[{\cal O}(x), {\cal O}(x')]=0,~ {\rm if}~ (x-x')^2<0
\eea
In the context of conformal field theories, these axioms are to be
appropriately interpreted and additional hypothesis
might be added.
 The Wightman
function $W_n(x_1,x_2,...x_n)$ is 
\bea
\label{sec2.2}
W_n(x_1,x_2,...x_n)=<0| \phi(x_1)\phi(x_2)...\phi(x_n)|0>
\eea
 We know that
these  vacuum expectation values are not ordinary functions but are
distributions. They are to be interpreted as linear
functionals as defined below
\bea
\label{sec2.3}
W[f]=\int d^4 x_1..d^4x_n W_n(x_1,x_2,...x_n)  f(x_1,x_2...x_n)
\eea
Thus we assign a complex number with the introduction of $f(x_1,x_2..x_n)$
 in the above the functional.
Moreover, $ f(x_1,x_2...x_n)$ are  infinitely differentiable functions. They 
 they vanish outside a bounded region of $4n$ dimensional space.
It is worth while to note that $W_n(f_1,f_2,...f_n)$ (indeed they are generally
denoted as functional $W[f]$) are well defined. The limit and continuity are
defined for such objects. We work with Wightman functions, 
$W_n(x_1,x_2,...x_n)$, in this article. We emphasize the fact that
these are distributions. Thus when we discuss convergence and boundedness etc.
of Wightman functions, they are to be understood in the sense that these are 
distributions, as defined above, with well behaved weight functions.
  Note that fields are operator valued
 distributions. Consequently, the spacetime
average of operators are interpreted as observables. Thus operators of the
 form
\bea
\label{sec2.4}
\phi[f]=\int d^4x \phi(x)f(x)
\eea
are meaningful.  It is assumed that $\phi[f]$
 is defined as the class of
 all infinitely differentiable functions of compact
support in spacetime. In this light, note that $W_{\phi}[f]=<0|\phi[f]|0>$
 is a linear
 functional with respect to $f$. More refined statements
can be made by introducing a sequence of test functions  $\{f_n(x) \}$ and by
specifying convergence properties \cite{book3}.
In the optics of the preceding discussions the n-point function
$W_n(x_1,x_2,...x_n)=<0| \phi(x_1)\phi(x_2)...\phi(x_n)|0>$
is a distribution in each of the variables $x_i$
\bea
\label{sec2.5}
W_n[f_1,f_2,....f_n]=<0|\phi(f_1)\phi(f_2).....\phi(f_n)|0>
\eea

 We  can give a precise interpretation to
$W_n(x_1,x_2,...x_n) $
in terms of the infinitely differentiable test functions $\{f_n \}$.
 Now on, when we deal with Wightman functions
$W_n(x_1,x_2,...x_n) $, it
is understood that they have interpretations as alluded to above.\\

As  a consequence of translational invariance of the theory;
  we conclude that
  $W_n(x_1,x_2,....x_n)$ depends on the difference of coordinates
\bea
\label{sec2.6}
W_n(x_1,x_2,....x_n)=W_n(y_1,y_2....y_{n-1})
\eea
where $y_i=x_i-x_{i+1}$. Moreover,  $W_n(y_1,y_2....y_{n-1})$ are
invariant under inhomogeneous Lorentz transformations; 
for a real  orthochronous  Lorentz transformation
\bea
\label{sec2.7}
W_n(y_1,y_2....y_{n-1})= W_n(\Lambda_r y_1,\Lambda_r y_2....\Lambda_r y_{n-1})
\eea
where $\Lambda_r$ is a real Lorentz transformations, ${\rm det}~\Lambda_r=1$.\\

{\it Local commutativity:} It follows  from axiom ($A4$) that  i.e.
$[\phi(x),\phi(x')]=0$ if $(x-x')^2<0$. Consequently,   
\bea
\label{sec2.8}
W_n(x_1,x_2,....x_j,x_{j+1},....x_n)=W_n(x_1,x_2,....x_{j+1},x_j,....x_n),~{\rm if}~ (x_j-x_{j+1})^2<0
\eea
Let ${\widetilde W}(p_1,p_2,...p_{n-1})$ denote the Fourier transform of
 $W_n(y_1,y_2....y_{n-1})$ then
\bea
\label{sec2.9}
W_n(y_1,y_2....y_{n-1})=\int d^4p_1d^4p_2...d^4p_{n-1}e^{-i\sum_{j=1}^{n-1}p_j.y_j}{\tilde W}_n(p_1,p_2...p_{n-1})
\eea
From the temperedness property we know that  $W_n$-functions  have
at most polynomial growth at infinity \footnote{see Froissart \cite{froissart}
for detailed arguments.}.  It is required that spectrum of
the physical states  must have timelike four momenta and positive energy.
 Therefore, from the stability of vacuum and support condition
\bea
\label{sec2.10}
{\widetilde W}_n(p_1,p_2...p_{n-1})=0,~~{\rm unless}~~p_i^2\ge 0,~p_i^0\ge 0,~
i=1,2...n-1.
\eea
The analyticity  structure of $W_n(x_1,x_2,...x_n)$,  in the coordinate space
 is related to the support
 properties of the Fourier transformed Wightman functions, ${\widetilde W}_n$,
  in
 the momentum space.\\  

The function ${\cal W}_n(\xi_1,\xi_2...\xi_{n-1})$ of complex variables
$\xi_j^{\mu}=y_j^{\mu}-i\eta_j^{\mu}, j=1,2,...n-1$
is defined as analytic continuation of the vacuum expectation values
$W_n(y_1,y_2....y_{n-1})$.
 The set of complex variables
$\{\xi_i^{\mu} \}$ are defined as follows: the real pair 
$\{ y_i^{\mu},\eta_i^{\mu}  \}$ are 
such that  $\eta_i^{\mu}\in V^+,~i.e.~ \eta_i^2\ge 0,~\eta_i^0\ge 0$.
 Thus $\{\eta_j \}$ is in the forward lightcone; moreover,
 $-\infty<y_{i\mu}<+\infty$. 
This is the definition of forward tube $T_{n-1}$. The primitive domain is
now  identified.   
Wightman functions,  the distributions,
  are boundary values of analytic functions i.e.
  \bea
  \label{sec2.10z}
  W_n(y_1,y_2...y_{n-1})={\rm lim}_{\{\eta_j\rightarrow 0 \}}
{\cal W}_n(\xi_1,\xi_2,...\xi_{n-1})
  \eea
Note that 
 ${\cal W}_n(\xi_1,\xi_2...\xi_{n-1})$ are also 
invariant under real orthochronous
Lorentz transformations; ${\rm det}~\Lambda_r=1$.
\bea
\label{sec29z}
{\cal W}_n(\xi_1,\xi_2...\xi_{n-1})=
{\cal W}_n(\Lambda_r\xi_1,\Lambda_r\xi_2...\Lambda_r\xi_{n-1})
\eea
$\Lambda_r$ is real  proper Lorentz transformation.   
Moreover, according to Hall and Wightman \cite{hw}, if 
${\cal W}_n(\xi_1,\xi_2...\xi_{n-1}) $ is analytic in the tube, $T_{n-1}$,
and is invariant under real orthochronous Lorentz transformations
 then ${\cal W}_n(\xi_1,\xi_2...\xi_{n-1}) $ is invariant under complex
Lorentz transformations where 
$(\xi_1,\xi_2,...\xi_{n-1})\rightarrow (\Lambda\xi_1,\Lambda\xi_2,...
\Lambda\xi_{n-1})$; note that  $(\xi_1,\xi_2,...\xi_{n-1})\in T_{n-1}$. Here 
$\Lambda\in SL_+(2{\bf C})$, ${\rm det}~\Lambda=1$. The set of 
points $(\Lambda\xi_1,\Lambda\xi_2,...\Lambda\xi_{n-1})$, for
arbitrary $\Lambda\in SL_+(2{\bf C})$,  define the extended 
tube $T_{n-1}'$. The enlargement of domain of holomorphy is achieved through
this procedure.
Important point to note is that $T_{n-1}$ does not contain
the real points of $\{\xi_j \}$. The extended tube contains real points
$\{y_i \}$. The axioms: Lorentz invariance, uniqueness of vacuum, stability
of vacuum and local commutativity lead to the following assertion.
The function ${\cal W}_n(\xi_1,\xi_2...\xi_{n-1}) $
exists and is analytic in the domain specified above. It is also single valued.
It is analytic in the domain which is union of the 
permuted  extended tubes.\\

The Wightman functions, $W_n(\{y_i \})$ are
  analytic functions of real variables $\{y_i \}$ 
when all of them are spacelike \cite{hw,ruelle}.
 The result of
 Hall and Wightman are very important
  for our purpose. We have noted earlier that when we define
${\cal W}_n(\{\xi_i \})$ in the extended tube they
  are analytic functions of complex four vectors $\{\xi_i^{\mu} \}$.
  Hall and Wightman  proved that
  if $W_n(\xi_1,\xi_2,...\xi_{n-1})$ is analytic in the four vector
variables $\{\xi_i^{\mu} \}$ which is
  invariant under the Lorentz transformations then it is an
analytic function of  scalar products of those complex four vectors.
 The
  statement, intuitively, seems to be reasonable; however, the proof
is quite formidable when
  mathematical rigor is enforced. There are two significant implications
of this theorem in the context of our work.
  First, the number of variables that the analytic function depends on
is reduced considerably. For example,
  the three point function, $W_3(x_1,x_2,x_3)$ depends on two four
  vectors, $y_1^{\mu}=(x_1-x_2)^{\mu}$ and $y_2^{\mu}=(x_2-x_3)^{\mu}$,
   from translation invariance of the theory. Thus it is a function of
  eight complex four vectors. In the light of Hall-Wightman theorem,
now we know that $W_3(\xi_1,\xi_2)$
  depends on the invariants $z_{ij}=\xi_i.\xi_j,i,j=1,2$. In order to
 bring out the essence of the second
  implication let us recall the following facts.  Notice that $z_{ij}$ is a
 complex symmetric matrix ($\xi_k=y_k-i\eta_k, k=1,2$)
  in the forward light cone and $\eta_j$ is in the interior of the lightcone.
  This set $\{z_{ij} \}$ is a domain of
  analyticity of an invariant function. Moreover, the Wightman function is
the boundary value when $\eta_j\rightarrow 0$.
  It has been proved by Hall and Wightman \cite{hw}
that there are set of points
$\{\xi_i \}$  on the
 boundary of the tube with following properties. These vectors can be used
to construct matrices $\xi_i.\xi_j$ which
  lie in the interior. Moreover, they have argued that an invariant
analytic function in the tube will not admit an
  arbitrary invariant distribution as boundary value. It has also been proved
that the boundary value is an analytic function
  of real variables $\xi_i.\xi_j, i,j=1,2,...n-1$, where $\eta_j=0$,
in a certain domain. The analytic function is
  uniquely determined once its values are known in some subdomain of the
boundary of the tube. These
  remarks  might look as if they are
  out of context to be interjected  at this juncture. The importance is
intimately related to the
  Jost theorem which we shall encounter in the discussion of crossing.\\

 The next question to ask
is where do the  real points of $T_{n-1}'$  reside and what are their
 attributes?
  Jost proved  the following: \cite{jost}
The real points  $\{y_1^{\mu},y_2^{\mu},...y_{n-1}^{\mu} \}$
 lie in the extended tube,
if and only if, the convex hull of   $\{y_1^{\mu},y_2^{\mu},...y_{n-1}^{\mu} \}$
only contains spacelike points.  To elaborate a little bit, the convex hull
of points $\{y_1^{\mu},y_2^{\mu},...y_{n-1}^{\mu} \}$ 
 is the set of all  four vectors
of the form 
$\lambda_1 y_1^{\mu}+\lambda_2 y_2^{\mu}+...\lambda_{n-1}y_{n-1}^{\mu}$
 where  the set $\{\lambda_i \}$ take positive
 real values, $\lambda_i\ge 0$ and
$\sum_i^{n-1}\lambda_i=1$. Therefore, the real points of the extended
tube are the ones for which if we take an arbitrary convex
linear combination $(\sum_{i=1}^{n-1}\lambda_i y_i^{\mu},~\lambda_i\ge 0, ~
\sum_i^{n-1}\lambda_i=1)$,  are  always spacelike i.e.
$(\sum_{i=1}^{n-1}\lambda_i y_i^{\mu})^2<0$, $\lambda_i\ge 0, ~
\sum_i^{n-1}\lambda_i=1)$. This is the Jost point. 
The following remarks are pertinent in order to appreciate the importance of 
the theorem of Jost.
We recognize that the determination of Jost point implies  the
existence of a domain consisting of points 
$y_{1\mu}^J,y_{2\mu}^J,..y_{n-1\mu}^J$ where $\{y_{k\mu}^J \}$ are real and
spacelike. These points lie in the interior of the extended tube $T_{n-1}'$.
However, they reside on the boundary of the forward tube, $T_{n-1}$. 
The Wightman function $W_n(y_1,y_2,...y_{n-1})$ is an analytic function of
the set of variables $\{y_{j\mu} \}$ at the Jost points. We can expand
$W_n$ in a convergent power series in these variables. We argue that if we know
the Wightman function in the neighborhood of 
$y_{1\mu}^J,y_{2\mu}^J...y_{n-1\mu}^J$, the Jost points, then the Wightman
function is uniquely determined for the set of variables 
$y_{1\mu},y_{2\mu},...y_{n-1\mu}$ of its argument. We shall utilize this fact
in the sequel.  
Moreover, their importance
is realized in the study of crossing and in the context PCT theorem.
 This is a very powerful result. We have been discussing crossing symmetry.
 We have alluded to the fact that the permuted Wightman
 functions coincide at spacelike point.
This theorem has far reaching consequences as would be evident in
the later part of this section as well as in the next section. \\

   The importance of Wightman formulation in conformal
field theories has been
recognized long ago  cite \cite{fgg1,polya2,luscher,mack1,mack2,fp2}.
Fradkin and Palchik \cite{fp2} have presented a very comprehensive
exposition in their book.
 Subsequently,
Mack {\cite{mack1}  has rigorously investigated the convergence of
operator product expansion in conformal field theories utilizing the
Wightman
formulation.
We have noted, in the last section, that not all conformal field theories
satisfy the spectrality condition of Wightman axioms. All the representations
 of
$SU(2,2)$, the covering group of the conformal group, have been classified
 quite sometime ago \cite{yao1,yao2,mack2,fp2}. According to the
classifications, a
conformal field belongs to unitary
irreducible representation of the covering group. The spectrality properties
have been  also investigated. Of importance, from our perspective, is a class
of conformal field theories known as nonderivative conformal field theories
which satisfy the Wightman axioms. They belong to the discrete
representation of the covering group.  Let us consider
the following situation:  $\{\phi^i \}$  be a st of 
conformal fields of dimensions $d_i $   and transform
as finite dimensional representations of the Lorentz group,
${\bf \cal L}=SL(2{\bf C})$.  Let $U\approx SU(2)$ the
rotation subgroup of ${\bf\cal L}$; $\bar U$ stands for set all finite
dimensional irreducible representations, denoted by $l$. Therefore, 
a representation, characterizing the field,  
would be collectively denoted by $\chi=(l, d)$.  As an illustration, consider 
the unitary irreducible representations of $SU(2,2)$,  labeled by (as noted)
 $l$,  which is a finite dimensional representation of the Lorentz 
group $SL(2{\bf C})$.
Suppose the highest weight representation of $l$ is $(2j_1,2j_2)$
(note that
$2j_1$ and $2j_2$ are nonnegative integers). Now, $\chi=(l,d)$ and
$d_{min}=2j_1+2j_2+2$ for nonzero $j_1$ and $j_2$. In addition the field
might
be endowed with internal quantum numbers. However, all along, we consider only
 a single Hermitian nonderivative conformal scalar field, $\phi$. The
Wightman axioms are respected by $\phi(x)$. 
We had discussed the structure of the Hilbert space and we recall that
it decomposes into several disjoint subspaces.
\bea
 \label{sec2.18}
 {\cal H}=\sum_{\chi}\oplus {\cal  H}^{\chi}
 \eea
 Here ${\cal H}^{\chi}$ stands for subspaces where each vector is
characterized by $\chi$ which are the quantum number belonging
 to the irreducible representations of the covering group alluded
to earlier. It follows from the algebra in OPE that even if we start with
a single field, $\phi(x)$, the operator product expansion of a pair
of this field needs infinite set of composite operator of nonderivative
type\cite{mack1}. Therefore, we need the states associated with them
when we construct the full Hilbert space, ${\cal H}$, which has subspaces
${\cal H}^{\chi}$. \\
Now  let us consider
 two four point Wightman functions for the conformal field theory:
 $W_4(x_1,x_2,x_3,x_4)=<0|\phi(x_1)\phi(x_2)\phi(x_3))\phi(x_4)|0>$ and
 $W_4(x_1,x_2,x_4,x_3)=<0|\phi(x_1)\phi(x_2)\phi(x_4)\phi(x_3)|0> $.
Where would they  coincide?
 We associate a tube (see the precise definition given earlier in this section)
 with respect to the first $W_4(x_1,x_2,x_3,x_4)$ and
another tube with the second $W_4(x_1,x_2,x_4,x_3)$.
 They are continued to one another
 as regular functions in the union of two extended tubes associated with each
of the functions. Note that the Jost point  has a real neighborhood
  in the extended tube.  Consider, for the case at hand,
$f_1(\xi_1,\xi_2,\xi_3)$ and $f_2(\xi_1',\xi_2',\xi_3')$ which are analytic
 in
 their corresponding extended tubes, $T_3'$. To remind the reader,
from translational invariance argument each 4-point
 Wightman function depends only on three coordinates, say $ y_1,y_2,y_3$ and
$y_1',y_2',y_3'$ and we can define corresponding
 extended tubes.  Let these two functions coincide for a real neighborhood in
 the extended tube
 \bea
 \label{sec2.14}
 f_1(y_1'',y_2'',y_3'')=   f_2(y_1'',y_2'',y_3'')
\eea
for $y_1'',y_2'',y_3''$  in a real neighborhood of a Jost point. The essential
conclusion of the Jost theorem is
\bea
\label{sec2.15}
f_1(\xi_1,\xi_2,\xi_3)= f_2(\xi_1,\xi_2,\xi_3)
\eea
This is  the edge-of-the-wedge theorem \cite{bot,t,epstein} 
and the implications
 of the theorem in the context of conformal bootstrap equation will be
discussed in Section 3.
This theorem was  first proved, in the context of
dispersion relations from LSZ axiomatic
field theory \cite{jost}.
We recapitulate a few points for the
sake of motivations and historical considerations .  
If we consider a four point scattering amplitude,
it has a right hand cut and a left hand cut. The right hand cut originates
 from the direct, $s$-channel, reaction; whereas, the left hand
cut arises from the crossed channel, the $u$-channel,  process. The
discontinuities across the cuts  are related
to the absorptive parts of the respective  amplitudes. The edge-of-the-wedge
theorem proves that the two
absorptive amplitudes are analytic continuation of each other
\cite{bot,t}.
Thus crossing was proved from analyticity
property of the amplitude and a dispersion relation could be written down.
We recall  that the theorem was proved for S-matrix elements
where external particles are on the mass shell and furthermore,
equations of motion were utilized in obtaining matrix elements of
source current commutators. The same arguments may be extended to the cases
of retarded and advanced commutators of currents.
In the context of conformal field theory,
we deal with Wightman functions. Therefore, a different route
has to be chosen when we intend to prove analyticity
and crossing in the present context.
The edge-of-the-wedge theorem has been proved for  Wightman functions 
by Epstein \cite{epstein}. Subsequently, 
the holomorphicity and envelop of holomorphy were studied by Streater
\cite{streaterjmp,streater1}  and Tomozawa \cite{tomozawa}.
 The preceding statements in the context of
(\ref{sec2.15})  is a qualitative and intuitive argument about the
edge-of-the-wedge theorem.
 We shall discuss this aspect in more details in the next section.
\\

\bigskip

\noindent 
{ \bf 2.2 Analyticity and Crossing Properties of Three Point Function}.\\

\bigskip

\noindent
Our objective is  to discuss analyticity and crossing properties of
the four point
Wightman function for Hermitian scalar nonderivative conformal field.
The three point Wightman function is
\bea
\label{sec2.20}
W_3(x_1,x_2,x_3)=<0|\phi(x_1)\phi(x_2)\phi(x_3)|0>
\eea
A crossed three  point function, for example,
 is $W_3(x_1,x_3,x_2)$ and the two
coincide when $(x_2-x_3)^2<0$. \\

We shall adopt the method employed
in \cite{mpla} to investigate analyticity and crossing properties 
of the three point function and report our further progress. 
There are six permuted 
three point functions. Our goal is to derive crossing relation
for a pair of permuted Wightman functions and  we address 
a simplified problem. We employ the analytic completion technique
for the problem at hand. We utilize the same prescription to
study analyticity and crossing for the four point function later in this
 section.
In what follows, we summarize the method employed by us  
to study the three point function.
 $W_3(x_1,x_2,x_3)$ depends on two
variables, $y_1=x_1-x_2$ and $y_2=x_2-x_3$, as a consequence of 
 translational invariance. The function of interest is
$W_3(x_1,x_2,x_3)-W_3(x_1,x_3,x_2)$. The recent progress made in this direction 
  by us is presented in the  sequel. The above  difference, 
$W_3(x_1,x_2,x_3)-W_3(x_1,x_3,x_2)$, may
be expressed as commutator 
$<0|\phi(x_1)[\phi(x_2),\phi(x_3)]|0>$ and it vanishes
when $(x_2-x_3)^2<0$. The goal is to study its analyticity
property which is intimately connected with crossing. We adopted a
variance of the representation, due to Dyson \cite{dyson111},
 of the double commutator of
three scalar fields:
$<0|[C(x_3),[B(x_2),A(x_1)]]|0>$ where $A,B,C$ are the three scalar fields.
 In fact Dyson's technique together with the arguments of
Streater \cite{streater} were suitably adopted by us to obtain a
representation for the VEV of our interest. Define
\bea
\label{sec2.21}
F(y_1,y_2)=\bigg( W_3(x_1,x_2,x_3)-W_3(x_1,x_3,x_2) \bigg)
\eea
Note that $F(y_1,y_2)=0$ for $y_2^2<0$ from microcausality. The Fourier
 transform, ${\tilde F}(p,q)$, of $F(y_1,y_2)$  has a representation
\cite{streater,mpla}
\bea
\label{sec2.22}
{\tilde F}(p,q)=\int {\bf \Psi}(p,u,s)\delta((u-q)^2-s^2)\epsilon ((u-q)_0)d^4uds^2
\eea
where $\epsilon((u-q)_0)$ is the Heaviside step function.
This is a generalized version of the Jost-Lehmann-Dyson
representation \cite{jl,dyson1}. The function ${\bf\Psi}(p,u,s)$ has following
properties:  It vanishes unless the hyperbola in the $q$-space i.e.
$(u-q)^2=s^2$ lies in the union of two domains characterized as
$q\in V^+\cup (p-q)\in V^+$. It was concluded that
${\bf\Psi}(p,u,s)=0$ except when the conditions
$p,u\in(u\in V^+\cap (p-u)\in V^+)$ are fulfilled.\\
It is necessary to identify the extended tubes for the two Wightman functions:
 (i)
 $W_3(\xi_1,\xi_2)$
 is regular in the extended tube $T'_2(\xi_1,\xi_2)$. (ii)  Similarly,
$W_3(\xi_1+\xi_2,-\xi_2)$ is regular in the extended tube
$T'_2(\xi_1+\xi_2,-\xi_2)$.
 If ${\widetilde W}_3(p,q)$ and ${\widetilde W}'_3(p,q)$ denote the
Fourier transforms of the Wightman functions $W_3(\xi_1,\xi_2)$ and
 $W_3(\xi_1+\xi_2,-\xi_2)$ respectively; note that the latter corresponds
 to the crossed 3-point Wightman function.
The point to note is that $W_3(\xi_1,\xi_2)=W_3(\xi_1+\xi_2, -\xi_2)$
in a domain where they are regular since these are Jost points i.e.
they corresponds to real points separated by spacelike distance. We conclude
 that they analytically continue to one another in the domain
\bea
\label{sec2.23}
{\bf T'}=T'_2(y_1,y_2)\cup T'_2(y_1+y_2,-y_2)
\eea
 Let us denote the Fourier transforms of $W_3(y_1,y_2)$ and
 $W_3(y_1+y_2,-y_2)$ respectively by ${\widetilde W}_3(p,q)$
 and ${\widetilde W}_3'(p,q)$.
 We can read off the support properties to be
  \bea
 \label{sec2.23}
 &&{\tilde W}_3(p,q)= 0,~~{\rm unless}~~ p^2>0,~p_0>0, ~~{\rm and}~~ q^2>0,~q_0>0 \nonumber\\&&
 {\tilde W}_3'(p,q)=0,~~{\rm unless}~~p^2>0,~p_0>0~~{\rm and}
 ~~(p-q)^2>0,~(p-q)_0>0
 \eea
Now  recall the expression for the three point function (\ref{conf10})
of the conformal field, $\phi(x)$.
Note that $y_1=(x_1-x_2),~ y_2=(x_2-x_3)$ and $x_1-x_3=y_1+y_2$. In terms
of $\xi_1$ and $\xi_2$, it assumes the form,
\bea
\label{sec2.23a}
{\cal W}_3(\xi_1,\xi_2)
= {\rm const.}\bigg[{{1\over{\xi_1^2\xi_2^2(\xi_1+\xi_2)^2}}}\bigg]
^{d/2}
\eea
Let us now consider the scale transformation: $\xi_ i\rightarrow\lambda\xi_i$.
Then 
\bea
\label{sec2.23b}
W_3(\xi_1,\xi_2)\rightarrow W_3(\lambda\xi_1,\lambda\xi_2)=&&
[({{1\over{\lambda}}})^3]^d
\bigg[{{1\over{\xi_1^2\xi_2^2(\xi_1+\xi_2)^2}}}\bigg]
^{d/2}\nonumber\\&&
=[({{1\over{\lambda}}})^3]^dW_3(\xi_1,\xi_2)
\eea 
We recall equations (\ref{conf10}) and (\ref{conf11}) for real spacelike
coordinate differences i.e. $(x_i-x_j)^2<0,~{\rm for}~ i,j=1,2,3$.  It is
quite transparent from the expression 
that the three point functions are analytic at the 
Jost point. 
Moreover, the denominator of (\ref{sec2.23a}) is analytic whenever, for 
real $\xi_i$
$\xi_1^2,{\rm and}~~\xi_2^2$ are negative and it is more transparent when we
consider $x_i$-variables.
Thus, expressed in terms of $x_i$'s   the two three
point functions $W3(x_1,x_2,x_3)$ and $W_3(x_1,x_3,x_2)$ coincide since both
can be considered as boundary values of corresponding analytic functions
at the Jost point. These two functions are analytic functions in the real
environment for spacelike separated points.
It follows from the Jost theorem and the edge-of-the-wedge theorem
that they are analytic continuation of each other. Thus the crossing is
established.\\ 

The preceding discussions lead to the conclusion that the two Wightman
 functions are boundary values of analytic functions with known support
properties.  Thus crossing is proved for a pair of Wightman functions
 such that one is obtained from the other from interchange of a pair of fields.
Let us  invoke the Hall-Wightman theorem \cite{hw} to discuss the analytic
 continuation. The three point function $W_3(x_1,x_2,x_3)$ depends of
 two variables
$y_i, i=1,2$. It is boundary value of an analytic function of two complex
 variables, $\xi_i, i=1,2$. Thus it depends on eight four-vectors.
 On the other
hand, it follows from the Hall-Wightman theorem that $W_3$ is a function
 of Lorentz invariant variables constructed from $\xi_1$ and $\xi_2$:
  it depends
on three complex variables: $z_{jk}=\xi^{\mu}_j\xi_{\mu k}$, $j,k=1,2$;
 expressed explicitly; $z_{11}=\xi_1^2,~z_2=\xi^2_2$ and $z_3=\xi_1.\xi_2$.
The Jost points are real and spacelike; moreover, they belong to $T_2'$.
 If the Jost points are denoted by $v_i,~ i=1,2$ that is $v_i^2<0$, and 
$v_1+\lambda v_2$ is also a Jost point with  $ 0<\lambda<1$.
   Thus for real $\xi_i,i=1,2$
\bea
\label{sec2.24}
(\xi_1+\lambda\xi_2)^2<0,~~{ \xi_1,\xi_2}~ {\rm real}
\eea
We derive a relationship among the variables $z_{jk}$.
Reality of $\lambda$ and condition (\ref{sec2.24}) imply
\bea
\label{sec2.25}
z_{12} >{\sqrt{z_{11}z_{22}}}
\eea 
  Obviously  the two Wightman functions
   $W_3(x_1,x_2,x_3)$ and $W_3(x_1,x_3,x_2)$  are
 equal at  the Jost point,  The corresponding extended tube for
 $W_3(x_1,x_3,x_2)$ is $T'_2(\xi_1+\xi_2, -\xi_2)$.
 Invoking Jost's theorem, we conclude
that the two Wightman functions are analytic continuation of each other.
  We have shown that the crossing holds for Wightman functions
while considered pairwise.  Therefore, all the six permuted
three point  Wightman
functions are analytic continuations of each other
when one  pair is considered
at a time. It is important to note, however,  that the envelope of
 holomorphy could be a much larger domain. We were contented to prove
crossing
in this simple approach.  \\

It is worth while to dwell upon the analyticity properties of the
three point function in the momentum space representation. 
The analyticity properties in momentum space representation of n-point
functions have been investigated in the past in the frameworks of axiomatic
field theories \cite{ab,ruelle61,araki61, vglaser}. The retarded functions, 
$R$-functions, defined in
the coordinate space are endowed with retardedness and causal properties, 
are defined to be 
\bea
R~\phi(x)\phi_1(x_1)...\phi_n(x_n)=&&(-1)^n\sum_P\theta(x_0-x_{10})
\theta(x_{10}-x_{20})...\theta(x_{n-10}-x_{n0})\nonumber\\&&
[[...[\phi(x),\phi_{i_1}(x_{i_1})],\phi_{i_2}(x_{i_2})]..],\phi_{i_n}(x_{i_n})]
\eea
with $R\phi(x)=\phi(x)$. Here P stands for all permutations $(i_1,...i_n)$ of
$1,2,...n$. The R-product is hermitial for hermitial fields $\phi_i(x_i)$ and
the product is symmetric under exchange of any fields
$\phi_1(x_1)...\phi_n(x_n)$. Notice that the field $\phi(x)$ is kept where it is
located in  its position.
We mention some of the important attributes of $R$-products.\\
(i) $R~\phi(x)\phi_1(x_1)...\phi_n(x_n) \ne 0$ only if
$x_0>~{\rm max}~\{x_{10},...x_{n0} \}$.\\
(ii) Another property of the R-product is that
\bea
R~\phi(x)\phi_1(x_1)...\phi_n(x_n) = 0
\eea
whenever the time component $x_0$, appearing in the argument of $\phi(x)$ whose
position is held fix, is less than time component of any of the four vectors
$(x_1,...x_n)$ appearing in the arguments of $\phi(x_1)...\phi(x_n)$.\\
(iii) Under Lorentz transformation 
\bea
\phi(x_i)\rightarrow \phi(\Lambda x_i)=U(\Lambda,0)\phi(x_i)U(\Lambda,0)^{-1}
\eea
 Therefore,
under Lorentz transformation $U(\Lambda,0)$
\bea
R~\phi(\Lambda x)\phi(\Lambda x_i)...\phi_n(\Lambda x_n)=U(\Lambda,0)
R~\phi(x)\phi_1(x_1)...\phi_n(x_n)U(\Lambda,0)^{-1}
\eea
And
\bea
 \phi_i(x_i)\rightarrow\phi_i(x_i+a)=e^{ia.P}\phi_i(x_i)e^{-ia.P}
\eea
 under spacetime translations. Consequently,
\bea
R~\phi( x+a)\phi( x_i+a)...\phi_n(x_n+a)=
e^{ia.P}R~\phi(x)\phi_1(x_1)...\phi_n(x_n)e^{-ia.P}
\eea
We conclude, therefore, that the vacuum expectation value of the R-product
dependents only on  difference between pair of coordinates: in other words it
depends on the
following set of coordinate differences:
$\xi_1=x_1-x,\xi_2=x_2-x_1...\xi_n=x_{n-1} -x_n$ as a consequence of
translational invariance.
It has been demonstrated that the Fourier transform of the VEV of
R-Products  are boundary values of
analytic functions of complexified momenta. There is a close
analogy between the three point Wightman function and the retarded three
point function from the analysis of Jost \cite{jost31} which has been further studied by Brown \cite{brown}.
\\

 We  study the analyticity properties of three point
function of conformal field theory in the momentum space.
Recently, there has been quite a bit of   interest in understanding  momentum
space descriptions of correlation functions  in conformal field
theories \cite{m1,m2,m3,m4,m5,m6,bg,mg,m7,m8,m9,m10,m11,m12,mpla}.
In particular, one of our interests is the study of the analyticity
properties of three point function in the momentum space
description \cite{bg,mg,mpla}. We were motivated
to undertake this study by the recent two papers \cite{bg,mg}.
We know, from Jost theorem,  that Wightman functions are analytic in the
spacelike regions i.e. when the coordinate separations are spacelike.
Moreover, the Fourier transform of a
Wightman function for spacelike coordinate differences would have conjugate
momenta lying in the spacelike region.
\\

Let us define the vacuum expectation value of $R$-product of $n$ scalar
field to be $r_n(y_1,y_2,...y_{n-1})$ where $y_i=x_i-x_{i+1}$. Therefore,
for a single type of  real scalar field, $\phi(x)$
\bea
\label{sec2.r3a}
r_n(y_1,y_2,..y_{n-1})=<0|R~\phi(x)\phi(x_1)...\phi(x_n)|0>
\eea
The Green function is
\bea
\label{sec2.3b}
G_n(p_1,p_2,...p_{n-1})=\int d^4y_1d^4y_2...d^4y_{n-1}
e^{i\sum_{j=1}^{n-1}p_j.y_j}r_n(y_1,y_2,...y_{n-1})
\eea
The next step is to define an analytic function of complexified
momentum variables: 
$p_j^{\mu}\rightarrow k_j^{\mu}=(p_j^{\mu}+iq_j^{\mu}), j=1,2,...n-1$.
The $(n-1)$ real four vectors $(p_j^{\mu},q_j^{\mu})$ are
such that $q_j\in V^+$ and $p_j^{\mu}$ is unrestricted.
The Fourier transform
\bea
\label{sec2.3c}
{\cal G}_n(k_1,k_2,..k_{n-1})=\int d^4y_1d^4y_2..d^4y_{n-1}
e^{i\sum_{j=1}^{n-1}k_j.y_j}r_n(y_1,y_2,...y_{n-1})
\eea
has good convergence property. Now the Green function $G_n(p_1,p_2,..p_{n-1})$
is boundary value of an analytic function
\bea
\label{sec2.3d}
G_n(p_1,p_2,..p_{n-1})={\rm lim}_{\{q_i\rightarrow 0 \}}
{\cal G}(k_1,k_2,...k_{n-1})
\eea
Ruelle \cite{ruelle61} has proved analog of Jost theorem in the momentum
space. Since $\{ q_i^{\mu} \}\in V^+$, we may choose a coordinate frame such
that $q_i^{\mu}=(q_i^0,{\bf 0})$ and there are no restrictions on $p^{\mu}_i$.
A simplified version of Ruelle's theorem can be expressed as follows:
 The function ${\cal G}(k^0_i, {\bf p}_i)$ can have singularities if $q^0_i=0$
and $p^{\mu}_i$ is {\it not} spacelike.
 This theorem is valuable for us in what follows.\\
The case of three point function will be taken up now in the light of
the preceding observations of Jost \cite{jost31}. We deal with the $R$-product
where the product is of  a single conformal scalar
field, $\phi(x)$ is defined to be
\bea
\label{sec2.39y}
R~\phi(x)\phi(x_1)\phi(x_2)=&&\theta(x_0-x_{10})\theta(x_{10}-x_{20})
[[\phi(x),\phi(x_1)],\phi(x_2)]\nonumber\\&& +\theta(x_0-x_{20})
\theta(x_{20}-x_{10})[[\phi(x),\phi(x_2)],\phi(x_1)]
\eea
Let us consider the implications of causality on the first double
commutator on the $r.h.s$ of (\ref{sec2.39y}), suppressing the presence of the
two $\theta$-functions,
\bea
\label{sec2.39s}
[[\phi(x),\phi(x_1)],\phi(x_2)]=0~{\rm for}~(x-x_1)^2<0,
~ {\rm or} ~[
(x-x_2)^2<0 ~ and~(x_1-x_2)^2<0]
\eea
Noting that $y_1=x-x_1$ and $y_2=x_1-x_2$, we may express these constraints
in terms of $y_i$-variables. Similar constraints follow for the second
terms of (\ref{sec2.39y}). We concentrate on
the  three point function $<0|\phi(x_1)\phi(x_2)\phi(x_3)|)>$ to study
momentum space analyticity properties. It
depends on two variables $y_1$ and $y_2$ and we introduce complexified
coordinates $\xi_1,\xi_2$. Moreover,
$W_3(\xi_1,\xi_2)=\int dp^4_1dp^4_2e^{-i(\xi_1.y_1+\xi_2.p_2)}
{\widetilde W}(p_1,p_2)$ and
 $p_1,p_2\in V^+$.
We also know from works of  Ruelle \cite{ruelle61} that,
in the coordinate space description,  we may go to a
frame where $\eta_i=(\eta^0_i,0,0,0), i=1,2$; this is permitted
 since $ \eta_i\in V^+$. Note   that
$-\infty< y_i<+\infty,i=1,2$, the
real part of $\xi_i$ (we remind the reader that
$\xi_j=y_j-i\eta_j, j=1,2$) . We also know that, for real $\xi_i$,
each of the three point functions ${\cal W}_3(\{\xi_i \}$ (appearing in the
$R$-product) are
analytic whenever $\{y_i \}\in T_2'$ and thus are spacelike,
from the Jost theorem.
We may adopt another theorem of Ruelle \cite{ruelle61} for this case;
${\cal{W}}_3(\eta^0_1,\eta^0_2, y_1,y_2)$  can only have singularities
if two of its arguments $(\eta^0_1,y_1)$ and $(\eta^0_2,y_2)$ are
such that $\eta^0_1=\eta^0_2$ and $y_1^{\mu}-y_2^{\mu} $ is {\it not} spacelike.
In what follows, we shall consider a three point Wightman
function $W_3(\xi_1,\xi_2)$ taking the clue from analysis of Jost \cite{jost31}.
  Let us consider the momentum
space representation of three point function
\bea
\label{sec2.39}
{\widetilde W}_3(p_1,p_2)=\int d^4y_1d^4y_2e^{i(p_1.y_1+p_2.y_2)}W(y_1,y_2)
\eea
Now if we want this Fourier transform to be convergent for real $y_1$ and 
$y_2$ then we complexify\footnote{ see the discussions of
Froissart \cite{froissart}.} the momentum variables $(p_1,p_2)$.
 As before,
 $(p_1,p_2)\rightarrow (k_1=p_1+iq_1,k_2=p_2+iq_2)$. Thus the tube,
${\widetilde T}_2$ is defined  in the momentum space and it is
 the primitive domain; moreover, the integral (\ref{sec2.39}) is convergent
since $q_1~{\rm and}~q_2\in V^+$.
The  momentum-space 3-point function is the boundary value of an analytic
function: ${\widetilde{\bf W}}(k_1,k_2)$
\bea
\label{sec2.40}
{\widetilde W}_3(p_1,p_2)={\rm lim}_{\{q_i \}\rightarrow 0 }
{\widetilde{\bf W}}_3(k_1,k_2)
\eea
 Assuming that Ruelle's theorem is valid for ${\widetilde{\bf W}}_3(k_1,k_2)$
we may invoke the momentum space theorem \cite{ruelle61}. Thus
${\rm lim}_{\{q_i \}\rightarrow 0 }
{\widetilde{\bf W}}_3(k_1,k_2)$ is analytic when real momenta $p_1^{\mu}$ and
$p_2^{\mu}$ lie in the spacelike region. We draw attention to a few points.
Let us consider the difference 
$R~\phi(x)\phi(x_1)\phi(x_2)-A~\phi(x)\phi(x_1)\phi(x_2)$ where the $A$-product
is defines such that $\theta((x_i-x_j)_0)$ is replaced by $\theta((x_j-x_i)_0)$.
Now consider the VEV of this difference and denote the VEV of $R$-product as 
$r(x,x_1,x_2)$
and correspondingly VEV of $A$-product as $a(x,x_1,x_2)$
\bea
\label{sec2.40a}
r(x,x_1,x_2)-a(x,x_1,x_2)=<0|[[\phi(x),\phi(x_1)],\phi(x_2)]|0>
\eea
We arrive at above expression using the properties of $\theta$-functions
and also the fact that $\theta(x_0)+\theta(-x_0)=1$.
This is the double commutator we have encountered before. If we open up
the double commutator it will be sum of four three point functions. We also 
know the constraints due to microcausality. Note that there are altogether 
six permuted Wightman functions and they are different. Let us consider the
momentum space functions. Then we complexify the momenta and define the
analytic functions in these complex variables. Now we appeal to Hall-Wightman
theorem \cite{hw} for these functions. Obviously, we have function which
depends on Lorentz invariants ${\tilde z}_{ij}= k_i.k_j, i,j=1,2,3$. We also
know $k_1^{\mu}+k_2^{\mu}+k_3^{\mu}=0$ due to
translational invariance and the complexified momenta are
 conjugate to $x,x_1,x_2$. The point to note is that now we can define a 
function which is analytic in the union of the extended tubes associated
with the three point analytic functions. If we go to boundary values in
each of the real valued momenta the corresponding Fourier transformed
Wightman functions are different. We see the power of analytic completion
in this simple example.
\\

It is intuitively  quite appealing. If a Wightman function is defined
for spacelike coordinate, in the coordinate space representation, the conjugate
momenta are also spacelike when we take the Fourier transform and study
analyticity. Our intuition guides that if those
coordinate points correspond to Jost point, we expect that the conjugate
momenta would be  spacelike. Moreover, the momentum space Green function
will be analytic in momentum variables in the real environment of these
Jost points (momentum space Jost points).   It is quite safe to conjecture
that our aforementioned arguments regarding analyticity of
${\widetilde{\bf W}}_3(k_1,k_2)$ are  valid. Let us start from the
momentum space Jost point where ${\widetilde{\bf W}}_3(k_1,k_2)$ is
analytic for ${\rm Re}~(k_1,k_2)$. Moreover, ${\widetilde{\bf W}}_3(k_1,k_2)$
is analytic in the momentum space extended tube, ${\tilde T}_2'$. If we
extend the argument of Jost, for momentum space three point function,
${\widetilde{\bf W}}_3(k_1,k_2)$, it is analytic for real spacelike $p_1,p_2$
in their neighborhood. Let us
  implement an infinitesimal
complex Lorentz transformation on these real momenta .
 The momenta $p_1,p_2$ will assume complex values
and arguments of ${\widetilde{\bf W}}_3(p_1,p_2)$ will be complex
(denote them as $k_1$ and $k_2$). The
resulting ${\widetilde{\bf W}}_3(k_1,k_2)$
would be analytically continued in ${\tilde T}_2'$.
 Therefore, the function is analytic in the domain belonging to ${\tilde T}_2'$.
We can say more. Note that $W_3(\xi_1,\xi_2)$ is a tempered distribution.
The Fourier transform of a tempered distribution is also a tempered
distribution  as proved by Froissart \cite{froissart}. 
Consequently, it is bounded as is defined
for a distribution i.e. the boundedness is to be understood in the sense
that  ${\widetilde{\bf W}}_3(k_1,k_2)$
is a distribution (see Froissart \cite{froissart} for detail discussion).   \\

We examine the implications of these general arguments in what follows.
 Recently, the properties of three point function
in the Lorentzian metric description has drawn  attentions
\cite{bg,mg}. Baurista and  Godazgar \cite{bg} have systematically
investigated the three point function in momentum space for the Euclidean
metric as well as for the Lorentzian signature metrics. They derive an
expression for the three point function as an integral  which
includes product of Bessel function as well as modified Bessel functions.
Their work revealed several interesting analyticity properties of
the Wightman function. Gillioz \cite{mg} studied
the properties of three point function in the momentum space with Lorentzian
signature metric. He employed  operator product expansion in momentum space
representation  and derived  Ward identities.
One advantage of this
approach is that he constructed states, utilizing ${\rm state}\leftrightarrow 
{\rm correspondence}$, in the momentum space in order to compute
three point function.
 He has demonstrated that the three point function
for scalars is expanded in a double hypergeometric series. Moreover,
he  explored its behavior in various kinematical limits. Our objective
is to study the analyticity properties in the momentum space in the light
of the preceding discussions.
\\

Let us consider the three point function given by (\ref{sec2.23a}). It is
bounded and analytic for spacelike $\{\xi_i \}, i=1,2$ \footnote{This 
boundedness property is to be understood as pertiment to distributions. 
Such a concept of boundedness is quite common in axiomatic S-matrix theory.
For example, when the notion of polynomial boundedness is mentioned for matrix
element of commutator of currents in the proof of Jost-Lehmann-Dyson theorem
theorem \cite{jl,dyson1} it is interpreted that the matrix element is
a distribution. Moreover, it fulfils polynomial boundedness so that the
integral requires only finite number of subtractions. Similar arguments are
advanced in proving the dispersion relations for 4-point amplitude where
absorptive parts are argued to be polynomially bounded.}.
  The Fourier
transform of the coordinate space three point function has been
investigated in detail in \cite{bg,mg}. Recall that the Fourier
transform of $W_3(x_1,x_2,x_3)$ would depend on three
momentum variable,
however, the total momentum conservation implies that only two of
the three momentum variables are independent.
 It follows from Lorentz invariance that
the three point function ${\widetilde W}_3(p_1,p_2,p_3)$ depends on the
 Lorentz invariants constructed from $p^{\mu}_i,i=1,2,3$. Furthermore,
conformal invariance imposes restrictions on its structure. We recall
the results of Gillioz \cite{mg} for our purpose.  The expression
for the three point function assumes the following generic form
(for $D=4$  in our metric convention)
\bea
\label{sec2.41}
<0|\phi(p_1)\phi(p_2)\phi(p_3)|0>= \theta(-p_1^0)\theta(p_3^0)
\delta^{(4)}(p_1+p_2+p_3)(p_2^2)^{(3d-8)/2}{\cal F}(\bigg({p_1^2\over{p_2^2}},
{p_3^2\over{p_2^2}} \bigg)
\eea
where the energy momentum conserving $\delta$-function implies
$p_1+p_2+p_3=0$. Moreover, ${\cal F}$ is a function of ratios of squares
of momenta.
Eventually, ${\cal F}$ gets  related to Appell's $F_4$ generalized
hypergeometric function of two variables and we refer to \cite{mg} for
details.\\

A few comments are in order at this stage: (i) Following Ruelle, we argue
that the momenta can be complexified as
 $p^{\mu}_i\rightarrow k^{\mu}_i=(p^{{\mu}}_i+iq^{\mu}_i)$ to define
an analytically continued three point function in the complex momentum
variables. We identify the primitive domain to be ${\tilde T}_2$ and
the corresponding extended tube as ${\tilde T}_2'$.  (ii) It follows from
momentum conservation that $p_2^2=(p_1+p_3)^2$ and the dimensionless
ratio of the momenta, appearing as arguments of the function ${\cal F}$ go
over to ${{p_1^2}\over{p_2^2}}\rightarrow {{p_1^2}\over{(p_1+p_3)^2}}$
and ${{p_3^2}\over{p_2^2}}\rightarrow {{p_3^2}\over{(p_1+p_3)^2}}$. The
function,
${\cal F}(\bigg({p_1^2\over{(p_1+p_3)^2}},{p_3^2\over{(p_1+p_3)^2}} \bigg)$
 is conformally invariant as is evident from the structure.
(iii) Now consider $p_1$ and $p_3$ to be spacelike.
 Furthermore, as $p_1^2$ and $p_3^2$ tend to asymptotic
values, with ${{p_1^2}\over{p_3^2}}\rightarrow{\rm constant}$,
 in the spacelike region, then ratios ${{p_1^2}\over{(p_1+p_3)^2}}$
and ${{p_3^2}\over{(p_1+p_3)^2}}$ tend to constants. Therefore, in this region,
the function,
${\cal F}(\bigg({p_1^2\over{(p_1+p_3)^2}},{p_3^2\over{(p_1+p_3)^2}} \bigg)$,
will tend to constant.
Moreover, we note that
the three point function is analytic for spacelike momenta and in its
a real neighborhood belonging to  ${\cal T}_2'$.
Furthermore, this function admits analytic continuation to the extended tube.
(iv) It is evident that the three point function depends on Lorentz invariant
variables and their ratios. This is analog of the Hall-Wightman theorem
\cite{hw} for three point function of our conformal field theory.
 The function of the
complex variables $k_i$ is a tempered distribution as the coordinate
space three point function is a tempered distribution.
\\

\bigskip

\noindent {\bf 2.3  Analyticity and Crossing for Four point Wightman Function}.
\bigskip

\noindent We proceed to study analyticity of four point Wightman function
following the prescription discussed above. The four point function
\bea
\label{sec2.26}
W_4(x_1,x_2,x_3,x_4)=<0|\phi(x_1)\phi(x_2)\phi(x_3)\phi(x_4)|0>
\eea
will be considered along with another permuted Wightman function.
 We shall proceed to investigate analyticity and crossing
adopting the spirit adopted for three point functions. 
The translational invariance implies that
 $W_4(x_1,x_2,x_3,x_4)=W_4(y_1,y_2,y_3)$ where $y_i=x_i-x_{i+1}$.
Let us recall the
expression for the four point function (\ref{conf7a}), (\ref{7b}) and
(\ref{7c}). In the light of the preceding discussion for the three point
function, we examine the analyticity of $W_4(x_1,x_2,x_3,x_4)$ for real
 values  of the coordinates $\xi_1,\xi_2,\xi_3$ and when the separations
are spacelike. Recall that four vectors, $y_i, i=1,2,3$,
 are all spacelike and their linear combinations with positive coefficients 
are also spacelike; a more precise statement is that for real $\xi_i$,
$\xi_i^2<0$ and $\sum_{i=1}^3(\lambda_i\xi_i^2)^2<0,~0<\lambda~{\rm}~
\sum_i\lambda_i=1$. It follows
from the definition of $Z_1$ and $Z_2$ that these 
are ratio of $y_i^{\mu}$'s and
the ratio is positive as long as each vector is spacelike. Moreover,
$Z_1$ and $Z_2$ are scale invariant. The denominator appearing as prefactor of
${\cal F}(Z_1,Z_2)$ is also positive. We invoke the Jost theorem and once
again, the edge-of-the-wedge theorem when we consider the two
four point functions $W_4(x_1,x_2,x_3,x_4)$ and $W_4(x_1,x_2,x_4,x_3)$ and
argue that they are analytic continuation of each other. Thus crossing
is demonstrated from this perspective.  \\

We intend to study the support properties from the Fourier transform 
of the four point function. We aim at deriving an integral representation
for the function, analogous to the Jost-Lehmann-Dyson representation as
our next step. Thus
\bea
\label{sec2.27}
{\widetilde W}_4(p_1,p_2,p_3)=\int d^4y_1d^4y_2d^4y_3e^{i\sum_{j=1}^3p_j.y_j}
W_4(y_1,y_2,y_3)
\eea
Note, from the spectral condition that
\bea
\label{sec2.28}
 {\widetilde W}_4(p_1,p_2,p_3)=0,~{\rm unless}  ~~ p_j\in V^+, ~j=1,2,3
 \eea
It is worth while to mention that the Fourier transform, 
${\widetilde W}_4(p_1,p_2,p_3) $, is also a distribution in the momentum 
variables.  Note that ${\widetilde W}_4(p_1,p_2,p_3) $ is a boundary value
of a holomorphic function as $W_4(y_1,y_2,y_3)$ is.
We argue that ${\cal W}_4(\xi_1,\xi_2,\xi_3),~{\rm recall} (\xi_i=y_i-i\eta_i)$
has a Laplace transform
\bea
\label{sec2.28a}
{\cal W}_4(\xi_1,\xi_2,\xi_3)=\int d^4p_1d^4p_2d^4p_3
e^{-i\sum_1^3p_j.(y_j-i\eta_j)}{\widetilde W}_4(p_1,p_2,p_3) 
\eea
It is holomorphic in the upper half-plane (note that $\eta_i\in V^+$ ensures
convergence). Moreover, the derivatives ${{dW_4(\{\xi_j \})\over{{d\xi_j}}}}$
exists and does not depend on direction. We remind that the function is
holomorphic in the forward tube and it is Laplace transform of a distribution.  
The distributions $ {\widetilde W}_4(p_1,p_2,p_3)$ vanish for 
$p_i^{\mu}\notin V^+$.
 The purpose of interjecting these remarks is to convey that
our next steps are based on the known analyticity properties of the
Fourier transformed four point function.\\
 We have introduced three complex variables and we have defined the
 forward tube.
Thus $W_4(y_1, y_2,y_3)$  is boundary value of analytic function where
 the complex coordinates
are in  $T_3(\xi_1,\xi_2,\xi_3)$. Next, the extended tube $T_3'$ is
obtained  by implementing orthochronous 
complex Lorentz transformations $SL_+(2{\bf C})$ on $\{\xi_j \}$.
In order
to proceed further, we define
the permuted  Wightman function where the locations of the two fields
 $\phi(x_3)$ and $\phi(x_4)$ are interchanged (recall the case of three
point function).
\bea
\label{sec2.28a}
W_4'(x_1,x_2,x_4,x_3)=<0|\phi(x_1)\phi(x_2)\phi(x_4)\phi(x_3)|0>
\eea
The new complexified
coordinates are $\xi_1'=\xi_1$, $ \xi_2'=\xi_2+\xi_3$, $\xi_3'=-\xi_3$,
expressed in terms of $\{\xi_j \}$.
The associated forward tube is
${\widetilde T}_3(\xi_1, \xi_2+\xi_3,-\xi_3)$. We can obtain the extended
tube
${\widetilde T}_3'$ once we have constructed
${\widetilde T}_3(\xi_1, \xi_2+\xi_3,-\xi_3)$.
The support properties of the
Fourier transform ${\widetilde W}_4' (p_1,p_2,p_3) $ of
$W_4'(x_1,x_3,x_2,x_4)$ are
\bea
\label{sec2.29}
{\widetilde W}_4' (p_1,p_2,p_3) \ne 0,~ {\rm for}~ p_1\in V^+,~ p_2\in V^+,~ 
p_2-p_3\in V^+,~{\rm and}~ p_3\notin V^+
\eea
which can be inferred from the expression for the Fourier transform.
 We now define a function
\bea
\label{sec2.30}
{\widetilde F}(p_1,p_2,p_3)=&&{\widetilde W}_4(p_1,p_2,p_3)-{\widetilde W}_4'(p_1,p_2,p_3)\nonumber\\ &&
={\widetilde W}_4\ne 0 ,~{\rm if}~p_1\in V^+,~p_2\in V^+,~p_3\in V^+, {\rm and} \nonumber\\
 &&
-{\widetilde W}_4'(p_1,p_2,p_3)\ne 0,~{\rm if}~ p_1\in V^+,~p_2\in V^+,~
p_2- p_3\in V^+,\nonumber\\ &&
~{\rm and}~p_3\notin V^+
\eea
We consider the two functions $W_4(\xi_1,\xi_2,\xi_3)$ 
and $W_4'(\xi_1,\xi_2+\xi_3,-\xi_3)$ to be analytic continuation
 of each other.
The corresponding domain is the union of two extended tubes
\bea
\label{sec2.31}
{\bf T'}=T_3'(\xi_1,\xi_2,\xi_3)\cup {\tilde T}_3'(\xi_1, \xi_2+\xi_3,-\xi_3)
\eea
We define the momentum space function
\bea
\label{sec2.31a}
{\widetilde F}(p_1,p_2,p_3)=\int <0|\phi(x_1)\phi(x_2)[\phi(x_3),\phi(x_4)]|0>
e^{i(p_1.y_1+p_2.y_2)}e^{+ip_3.y_3}d^4y_1d^4y_2d^4y_3
\eea
Notice that, from microcausality,
\bea
\label{sec2.31b}
 \int d^4p_3e^{-ip_3.y_3}{\widetilde F}(p_1,p_2,p_3)=0~{\rm if}~y_3^2<0
\eea
We may use similar arguments as we used for the 3-point function to
introduce an integral representation for ${\widetilde F}(p_1,p_2,p_3)$,
following the arguments of \cite{streater1}
\bea
\label{sec2.31c}
{\widetilde F}(p_1,p_2,p_3)=&&\int {\bar\Phi}(u,p_2,v,s,\kappa)
\delta[(p_1-u)^2-s]
\epsilon([(p_1-u]_0)\nonumber \\&&
\delta[(p_3-v)^2-\kappa]\epsilon([p_3-v]_0)     
d^4ud^4vdsdk
\eea
The function $ {\bar\Phi}(u,p_2,v,s,\kappa)$ satisfies following properties:
(i)    
Note that
\bea
\label{sec2.31d}
{\bar\Phi}(u,p_2,v,s,k)=0,~{\rm unless}~(p_1-u)^2>0,p_2^2>0,
 (p_3-v)^2>0
\eea
(ii) The condition $(p_1-u)^2=s$ and $(p_3-v)^2=k$ defines a pair of
hyperboloids with four momenta
satisfying above constraints. The function 
${\bar\Phi}(u,p_2,v,s,k)$ vanishes unless these conditions are fulfilled.
This is analog of the Jost-Lehmann-Dyson
representation and was considered by Streater \cite{streater1,streaterjmp}.\\
Remark: We have considered two permuted Wightman functions and it is
argued that their domain of holomorphy is the union  two extended
tube $T_3'(\xi_1,\xi_2,\xi_3)$ and
${\widetilde T}_3(\xi_1,\xi_2+\xi_3,-\xi_3)'$.
Let us construct the Lorentz invariant variables from the set $\{\xi_i \}$
for the application of Hall-Wightman theorem.
$z_{ij}=\xi_i.\xi_j, i,j=1,2,3$. We remind the reader that the three point
Wightman function has six permutations of the ordering of
the field. Similarly, the four point function has twelve permutations.
 The determination of the envelope of holomorphy
for this case is a formidable task. The problem for this case has not
been analyzed in so much of  details as has been done for
three point functions by K\"alle\'n and Wightman \cite{kw}. We recall that
 Lorentz invariance and local commutativity  are the two principles which are
instrumental to study the general structures. In the case of three point
function  (the VEV ) can be continued to
a function regular in a certain domain ${\cal M}_3$ in the space of scalar
products of $\{\xi_i \}$. We have seen that local commutativity
plays an important role to identify the union of domains which are formed
when we consider permuted field configurations.  There have
been several attempts   \cite{kt,kallen,book1} to implement 
  similar prescription for the four
point function and identify the domain ${\cal M}_4$ and they
have achieved some  success; however, the detail analysis
carried out for the three point function \cite{kw} has not been accomplished
for the four point function.
We shall be contented with the analysis for a pair of functions as envisaged
 above. We shall consider  a simplified situation
guided by our previous attempt \cite{mpla}. We
consider two spacelike vectors on ${\cal M}_3$ which is a subspace of
${\cal M}_4$. A third spacelike vector is considered which lies
in ${\cal M}_4$ but not in ${\cal M}_3$.  Then we determine the Jost point.
This procedure enables us to determine points of
${\cal M}_4$ similar to the condition derived for three point functions.
 Now there will be more inequalities (see \ref{sec2.25})
compared to the case of $W_3(\xi_1,\xi_2)$.  \\

We begin with the set of complexified four vectors $\{\xi^{\mu}_i \}, i=1,2,3$
 which are defined in manifold ${\cal M}_4$. Let us consider
a submanifold ${\cal M}_3$ in ${\cal M}_4$. Consider, two spacelike
real vectors which lie in ${\cal M}_3$. Their linear combination,
with a positive real coefficient, bounded by unity,
 is also a spacelike vector. Now choose
a spacelike vector in the compliment of ${\cal M}_3$. We should
be in a position to obtain constraints on these three real spacelike
vectors which would be similar to (\ref{sec2.25}).  For the case at hand,
we have to consider the decomposition of ${\cal M}_4$ into
 ${\cal M}_3\oplus{\cal M}$; ${\cal M}$ lies in the space compliment
to ${\cal M}_3$,
for three different configurations as discussed below.
\\
We go through the following steps:\\
(i) Let $\xi_1$ be a real spacelike vector in ${\cal M}$ which
is compliment of corresponding ${\cal M}_3$ i.e. ${\cal M}_4={\cal M}_3\oplus
{\cal M}$.\\
(ii) $\xi_2$ and $\xi_3$ are two spacelike vectors which are lying in
${\cal M}_3$. Thus $\xi_2+\lambda\xi_3,~ 0<\lambda<1$ is 
also spacelike where $\lambda$ is positive  and real.
 Let us consider the  properties  of the three vectors in three different
configurations.\\
 Case (a): 
Recall that $\xi_1^2<0$ and $\xi_1^2(\xi_2+\lambda\xi_3)^2>0$ and therefore,
${\sqrt{[\xi_1^2(\xi_2+\lambda\xi_3)^2}]}>0$ with $\lambda>0$. We
conclude that that
\bea
\label{sec2.33}
 [\xi_1.(\xi_2+\lambda\xi_3)]^2< \xi_1^2(\xi_2+\lambda\xi_3)^2
\eea
Note that the $r.h.s$ of the above equation is positive.
As before, define the Lorentz invariant  Hall-Wightman variables
 $z_{ij}=\xi_i.\xi_j,~i,j=1,2,3$.  We choose $\{\xi_i \}$ to be real in
 the light of
the preceding discussion to derive relations among $z_{ij}$ from the
 constraint that $\lambda>0$. When expressed in terms of $z_{ij}$,
The inequality  (\ref{sec2.33}) translate to
\bea
\label{sec2.34}
z_{11}z_{22}-z_{12}^2+2\lambda(z_{11}z_{23}-z_{12}z_{13})+\lambda^2(z_{11}z_{33}-z_{13}^2)>0
\eea
We can derive an inequality, similar to (\ref{sec2.25}) if we demand
 that $\lambda$ be positive and real (recall $0<\lambda<1$ here).
  Instead of obtaining
such constraints in a case by case basis let us consider the  other two cases.
We choose two spacelike vectors
in a subspace ${\cal M}_3$ and another one which lies in its compliment.
  The  other two case are\\
 (b) $\xi_1$ and $\xi_3$ lie in an
 ${\cal M}_3$ and $\xi_2$ in its compliment, ${\cal M}$. \\
(c) Here we have another permuted scenario
 for $\{\xi_1, \xi_2,\xi_3 \}$.\\

We shall obtain two more equations from the case (b) and (c) which will
 be analogous to (\ref{sec2.34}).
We can derived the set of constraints in a more efficient and elegant manner.
  Define the matrix
\bea
\label{sec2.35}
{\bf Z}=\pmatrix{z_{11} & z_{12} & z_{13} \cr
z_{21} & z_{22} & z_{23} \cr
z_{31} & z_{32} & z_{33} \cr }
\eea
Note that  $\bf Z$-matrix is symmetric, ${\bf Z}^T={\bf Z}$ as was introduced
 by Hall and Wightman \cite{hw}.
 Let us define a matrix ${\widetilde{\bf M}}$
\bea
\label{sec2.36}
({\widetilde{\bf M}})_{ij}=({\rm det~{\bf Z}})({\bf Z}^{-1})_{ij}
\eea
The constraint equations arising from the requirement that $\lambda >0$ is
 expressed in terms of the elements
of the ${\widetilde{\bf M}}$. Notice that we have to solve for  the analog of
 equations like  (\ref{sec2.34})  to
derive the requisite constraints in terms of the matrix elements of
 ${\widetilde{\bf M}}$.
Moreover, there will be altogether three equations. The conditions
 are given below:\\
($a'$) We start with the configurations $\xi_1^2<0$ and 
$(\xi_2+\lambda\xi_3)^2<0, 0<\lambda<1$ where $\xi_2$ and $\xi_3$ are spacelike.
Positivity of $\lambda$ leads to following inequality:
${\widetilde{\bf M}}_{23}^2>{\widetilde{\bf M}}_{33}{\widetilde{\bf M}}_{22}$.
\\
($b'$) For the case $\xi_2$, $\xi_3$ and $\xi_1$ spacelike
with $(\xi_1+\lambda\xi_3)^2<0, 0<\lambda<1$, 
the correspondingly the condition is :
${\widetilde{\bf M}}_{13}^2>{\widetilde{\bf M}}_{33}{\widetilde{\bf M}}_{11}$.
\\
($c'$)  For the third case i.e.  $\xi_3$ real and spacelike and two real
spacelike
vectors $\xi_1$ and $\xi_2$ lying in a a complement. Thus
$(\xi_1+\lambda\xi_2)^2<0$ for $0<\lambda<1$.
The condition becomes:
${\widetilde{\bf M}}_{21}^2>{\widetilde{\bf M}}_{11}{\widetilde{\bf M}}_{22}$.\\
Notice that same procedure can be adopted to derive constraints for the
real values of the  the complexified variables, $\xi_1',\xi_2',\xi_3'$;
($\xi_1'=\xi_1+\xi_2$, $ \xi_2'=-\xi_2$, $\xi_3'=\xi_2+\xi_3$).
Moreover, $W_4(x_1,x_2,x_3,x_4)$ and
$W_4'(x_1,x_3,x_2,x_4)$
are equal  for $(x_2-x_3)^2<0$ which is a Jost point and corresponds to
 $({\rm Re}~\xi_2)^2<0$. We have proved crossing for the
pair of Wightman functions and identified the analyticity regions.\\

Let us recall the structure of the four point function for the
conformal field theory, expressed in terms of the real $xi_i$-variables
\bea
\label{sec2.38}
W_4(\xi_1,\xi_2,\xi_3)=\bigg[{{1\over{\xi_1^2\xi_3^2}}}\bigg]^d
{\tilde{\cal F}}({\tilde Z}_1,{\tilde Z}_2)
\eea
Now ${\tilde Z}_1$ and ${\tilde Z}_2$ are cross ratios
of variables $\{\xi_i \}$.
\bea
\label{sec2.38a}
{\tilde Z}_1= {{\xi_1^2\xi_3^2}\over{(\xi_1+\xi_2)^2(\xi_2+\xi_3)^2}}
\eea
and
\bea
\label{sec2.38b}
{\tilde Z}_2= {{(\xi_1+\xi_2+\xi_3)^2\xi_2^2}\over{(\xi_1+\xi_3)^2(\xi_2+\xi_3)^2}}
\eea
Notice also that the cross ratios are conformally invariant. The prefactor
of the function of cross ratio transforms according to the known rule. Thus
\bea
\label{sec2.38c}
W_4(\lambda\xi_1,
\lambda\xi_2,\lambda\xi_3)
=(\lambda)^{-4d}W_3(\xi_1,\xi_2,\xi_3)
\eea
If we now recall equations (\ref{conf7a}), (\ref{7b}) and (\ref{7c})
\bea
 \label{conf7x}
 W_4(x_1,x_2,x_3,x_4)=&&<0|\phi(x_1)\phi(x_2)\phi(x_3)\phi(x_4)|0>\nonumber\\&&=[{1\over{({\tilde{z}}_{12}^2-i\epsilon{\tilde z}_{12}^0)
 ({\tilde{z }_{34}^2-i\epsilon}{\tilde z}_{34}^0})}]^{d}{\cal F}(Z_1,Z_2)
 \eea
 with 
 ${\tilde z}_{12}=x_1-x_2, {\tilde z}_{34}=x_3-x_4, {\bar z}_{13}=
x_1-x_3, {\tilde z}_{24}=x_2-x_4, {\tilde z}_{14}=x_1-x_4, 
{\tilde z}_{23}=x_2-x_3$.
\bea
 \label{7y}
 Z_1={{({\tilde z}_{12}^2-i\epsilon{\tilde z}_{12}^0)({\bar z}_{34}^2-i\epsilon{\tilde z}_{34}^0)}\over{({\tilde z}_{13}^2-i\epsilon{\tilde z}_{13}^0)
 ({\tilde z}_{24}^2-i\epsilon {\tilde z}_{24}^0)}}
 \eea
 and
 \bea
 \label{7z}
 Z_2={{({\tilde z}_{14}^2-i\epsilon{\tilde z}_{14}^0)
({\tilde z}_{23}^2-i\epsilon{\tilde z}_{23}^0)}
\over{({\tilde z}_{13}^2-i\epsilon{\tilde z}_{13}^0)
 ({\tilde z}_{24}^2-i\epsilon {\tilde z}_{24}^0)}}
 \eea
  $\cal F$ is a function which depends on cross ratios  and its
form is determined by the model under considerations.
Now we consider real  $x_i$ such that $(x_i-x_j)^2<0~i,j=1,2,3,4$ 
(without $i\epsilon$ factors) then this is a Jost point.
The four point function is analytic at the Jost point. Moreover,
the permuted four point function will coincide with (\ref{conf7x}) at the
Jost point it is clear that the four point function
 Jost point.
 Therefore, the crossing will be valid in this region.
We are not in a position to make more accurate statements since the
functional dependence of cross ratios are not known unless we appeal
a specific model.\\

\bigskip

\noindent{\bf 3. Analyticity and Causality in Operator Product
Expansions and Conformal Bootstrap
Equation}.
\bigskip

\noindent In this section we discuss analyticity and causality
properties of the four point Wightman functions in the context
of conformal bootstrap equation. First, we focus attention on
operator product expansion of a pair of conformal fields. The
proposal of Otterson and Zimmermann \cite{oz} is appropriately
modified in the context of CFT. Their investigation is based on
the work of Wilson and Zimmermann \cite{wz} who rigorously
studied operator expansion in QFT from the Wightman axioms.
The analyticity property of the matrix elements of composite operators
appearing in OPE are investigated. Mack \cite{mack1} had initiated
the study of the properties OPE in CFT from Wightman axiom view point.
When we invoke microcausality for the commutator of two operator
product expansions, the matrix element of the (difference) of two
composite operators are constrained. Therefore, the analog
of the Jost-Lehmann-Dyson representation can be derived.\\

The second part of this section is devoted to derive bootstrap equation
in a novel way through the PCT theorem. The equivalence between PCT
theorem and weak local commutativity (WLC) is invoked to related
two four point functions. The conformal partial wave expansion is
implemented on each the four point Wightman functions to derive the
bootstrap equation. We draw attention to following points to
highlight our approach. The first point to note is that the two four point
functions coincide at the Jost point. 
 It should be noted that the corresponding
Fourier transformed Wightman functions depend on conjugate momenta
which are spacelike and we term them as 'unphysical'.
Moreover, each four point function is boundary value of an analytic function.
These Wightman functions are 
 analytic in real neighborhood of Jost point. They are analytic
in corresponding extended tubes. 
Thus it will be demonstrated, with chain of arguments,
that the two four point functions are analytic continuation of each other.
The second point is the following. The conformal bootstrap
equation hold at real points where coordinates have spacelike separations.
As will be shown later, the bootstrap equations can be interpreted
as boundary values of analytic functions defined over extended tubes.
We feel that this proof is quite novel in the sense that the power
of PCT theorem and its equivalence with WLC enables us to derive
the bootstrap equation in CFT rigorously.
\\
\bigskip

\noindent {\bf 3.1. Operator Product Expansion, Causality and Analyticity}.
\bigskip

\noindent
Let us recapitulate the Wilson's operator product  expansion
proposal  for
the product of two scalar fields $A_1(x_1)A_2(x_2)$ as envisaged in \cite{oz},
based on the works of Wilson and Zimmermann \cite{wz}.
In the case of scalar fields 
\bea
\label{sec3.1}
A_1(x_1)A_2(x_2)=\sum_{j=1}^kf_j(\rho)C_j(x,\zeta,\rho)+{\bf R}(x,\zeta,\rho)
\eea
where $x_1^{\mu}=x^{\mu}+\rho\zeta^{\mu},~ x_2^{\mu}
=x^{\mu}-\rho\zeta^{\mu},~\rho>0$ and
 ${\bf R}(x,\zeta,\rho)$ stands for the remainder of the series
collectively. 
The series can be so organized that coefficients
would  satisfy
\bea
\label{sec3.2}
{\rm lim}_{\rho\rightarrow 0}~{{f_{j+1}(\rho)}\over{f_j(\rho})}
=0,~~~{\rm lim}_{\rho\rightarrow 0}~{{\bf R}\over{f_j(\rho)}}=0
\eea
  Note that $\{f_j(\rho) \}$ are C-numbers and they become
singular as $\rho\rightarrow 0$.
The operators, $C_j$, depend on the vector $x^{\mu}$
which is identified as the center of mass point. It also depends
 on another vector $\zeta^{\mu}$ which  is proportional to
the distance between
the two operators $A(x_1)$ and $A_2(x_2)$. We note that the
 $\zeta$-dependence is connected with the directional
dependence of the set of operators $\{ C_j \}$.  It becomes obvious
 from the relations
\bea
\label{sec3.3}
\zeta^{\mu}={{\kappa^{\mu}}\over{\sqrt {\kappa^2}}},~~
\rho={\sqrt {\kappa^2}}
\eea
We can reexpress (\ref{sec3.1}) as
\bea
\label{sec3.4}
A_1(x+\kappa)A_2(x-\kappa)=\sum_{j=1}^kf_j
({\sqrt{\kappa^2}})C_j(x,{{\kappa}\over{\sqrt{\kappa^2}}})+{\bf R}
\eea
The scalar fields, $A_(x_1)$ and $A_2(x_2)$ respect the Wightman axioms.
  The  operators $C_j(x,\zeta) $ are local in $x$
for a given $\zeta$.
 Otterson and Zimmermann \cite{oz} have  investigated the
relationship between causality and analyticity rigorously for OPE
of two scalar fields.\\

 Our intent is to investigate relationship between causality
 and analyticity in conformal field theory
 in the frameworks of Wightman axioms and adopt  the
 formalism introduced by \cite{oz}. We remind  that not all conformal
 field theories fulfill the requirements of Wightman axioms. The
 Hermitian scalar nonderivative field, $\phi(x)$,
  respects Wightman axioms. Its Fourier transform, ${\tilde\phi}(p)$,
 satisfies the spectrality condition i.e.
 $p\in V^+$.  We consider a single conformal scalar field, $\phi(x)$.
 The operator product expansion is
 \bea
 \label{sec3.5}
 \phi(x_1)\phi(x_2)=\sum_{\chi}\sum_jf_j^{\chi}(\rho)C_j^{\chi}(x,\zeta)
 \eea
 Mack \cite{mack1} has investigated the convergence of the operator
product expansion rigorously for nonderivative scalar conformal field theory.
   The coefficients $\{f^{\chi}_j(\rho) \}$ are  $C$-number
functions which become singular as $\rho\rightarrow 0$.
 For the nonderivative field, $\phi(x)$, in the OPE,  the complete set
of local operators, $\{ C^{\chi}_j \}$, are also nonderivative operators
\cite{mack1}.
It is to be understood that in the operator product expansion there
might be derivative of field satisfying the desired properties. We mention that
Mack \cite{mack1} considered operator product expansion for
 nonderivative field and convergence property of a matrix
element of the type 
$<\psi(f)|\phi(x/2)\phi(-x/2)|0>$. Here $|\psi(f)>$ is
a vector in the Hilbert space, ${\cal H}$, defined to be
\bea
\label{sec3.5a}
|\psi(f)>=
\sum_l\int dx_1dx_2,...dx_lf_l(x_1,x_2,...x_l)\phi(x_1)\phi(x_2)
...\phi(x_l)|0>
\eea
 Note that here one considers a single nonderivative field
$\phi(x)$ and $f(x_1.x_2,..x_l)$ is the usual weight function.  
He presented the matrix element of the operator product of two fields. In
eq. (\ref{sec3.5}), 
the sum over  $\chi$ goes  over all unitary irreducible finite dimensional
 representations of the covering group of the
 conformal group, $SO(4,2)$, usually identified as $SU(2,2)$.  We
adopt a notation $\chi=[l,\delta]$; $l$ corresponds to finite dimensional
 irreducible representation of $SL(2, {\bf C})$; it is the 'Lorentz spin'
 and $\delta\ge \delta_{min}$, real dimension. 
  OPE, as noted in section 1,  requires
 infinite set of operators belonging to irreducible representations
of the covering group. Therefore, a Hilbert space is associated with each of
 these local fields characterized by $\chi$. Thus the full Hilbert space,
 ${\cal H}$, is decomposed as $ {\cal H}=\sum_{\chi}\oplus {\cal  H}^{\chi}$
 as discussed already. 
 It will suffice to consider a series expansion
 for a given $\chi$ (i.e. fixed $\chi$)
 in order to investigate causality and analyticity
properties. We pick up a generic term in the double series expansion
(\ref{sec3.5}) and define
 \bea
 \label{sec3.8}
 \sum_j f^{\chi}_j(\rho)C^{\chi}_j(x,\zeta)=
\sum_j{\tilde f}_j{\tilde C}_j(x,\zeta),~ {\rm for~a~given~}\chi
\eea
Note that the $r.h.s$ of the  OPE, (\ref{sec3.5}), is a double sum.
It suffices for us to analyze properties of (\ref{sec3.5}) in the present
context. Therefore, 
we investigate the causality and
analyticity properties of a single sum over $j$ in  (\ref{sec3.8}) i.e.
look at the summed over $j$ term for a given $\chi$.
 The conclusion drawn from here will hold for each of 
 term of (\ref{sec3.5}) in the sum over $\chi$ of the double sum.
  Here ${\tilde f}_j(\rho)~{\rm and}~{\tilde C}_j(x,\zeta)$
stand for  $f^{\chi}_j(\rho)~{\rm and}~C^{\chi}_j(x,\zeta)$ respectively
so that we do not
carry the index $\chi$ everywhere. The series is so organized,
as was adopted in \cite{oz},
for a given sector of $\chi$,  that coefficients of the operator
${\tilde C}_j(x,\zeta)$
satisfy a condition analogous to (\ref{sec3.2}); i.e.
\bea
\label{sec3.9}
{\rm lim}_{\rho\rightarrow 0}
~{{{\tilde f}_{j+1}(\rho)}\over{{\tilde f}_j(\rho})}=0
\eea
We define
\bea
\label{sec3.10}
P_j(x,\zeta,\rho)={\tilde f}_j(\rho){\tilde C}_j(x,\zeta)
\eea
following the clue from Otterson and Zimmermann \cite{oz}. Consequently,
\bea
\label{sec3.11}
{\tilde C}_j(x,\zeta)=
{\rm lim}_{\rho\rightarrow 0}{{P_j(x,\zeta,\rho)}\over{{\tilde f}_j(\rho)}}
\eea
Define Fourier transform of ${\tilde C}_j(x,\zeta)$ as
\bea
\label{sec3.12}
{\bar C}_j(x,u)=
{{1}\over {2\pi^2}}\int d^4\zeta e^{i\zeta.u}{\tilde C}_j(x,\zeta)
\eea
Let us envisage two state vectors $|p>~{\rm and }~|q>$ in the Hilbert
space ${\cal H}^{\chi}$ for a given $\chi$ in the momentum space
representation. Thus $P_{\mu}|p>=p_{\mu}|p>$ and  $P_{\mu}|q>=q_{\mu}|q>$.
Now  consider the matrix element which satisfies
\bea
\label{sec3.13}
<p|{\bar C}_k(x,u)|q>=0,~{\rm unless}~u\in V^+
\eea
In fact if $<{\cal \phi}(p)| ~{\rm and }~|\psi(q)>$ are
arbitrary states which are superposition of momentum states the matrix
element $<{\cal \phi}(p)|{\bar C}_k(x,u)|\psi(q)> =0,~{\rm unless}~u\in V^+$.
We recall
that Wilson and Zimmermann \cite{wz} introduced additional hypothesis
when they envisaged operator product expansion in the Wightman formulation
for QFT. The same hypothesis was invoked in \cite{oz}. In the present case,
we also assumes that similar properties are satisfied by the
matrix elements of operator product expansion of two conformal
fields taken between two momentum space state $|p>$ and $|q>$ obtained
from state operator correspondence with ${\tilde \phi}(p)$ and ${\tilde \phi}
(q)$. Moreover, ${\tilde \phi}(p)$ and ${\tilde \phi}(q)$ satisfy the 
spectrality condition i.e. $p\in V^+$ and $q\in V^+$.
The microcausality constraint is
\bea
\label{sec3.14}
[\phi(x_1),\phi(x_2)]=0,~{\rm if}~ (x_1-x_2)^2<0
\eea
This condition, for the local composite operators in the OPE, translates to
\bea
\label{sec3.15}
{\tilde C}_j(x,\zeta)={\tilde C}_j(x,-\zeta),~{\rm for}~\zeta^2<0
\eea
Thus the function
\bea
\label{sec3.16}
{\cal F}_j(x,\zeta)=
<{\cal\phi}(p)|({\tilde C}_j(x,\zeta)-{\tilde C}_j(x,-\zeta))|\psi(q)>=0,
~{\rm for}~\zeta^2<0
\eea
In the Fourier space,
\bea
\label{sec3.17}
{\tilde {\cal F}}_j(x,u)=<{\cal \phi}(p)|({\bar C}_j(x,u)-
{\bar C}_j(x,-u))|\psi(q)>=0,~{\rm if}~u^2<0
\eea
Notice that ${\tilde{\cal F}}_j(u)$ satisfies all the conditions of
Jost-Lehmann-Dyson theorem \cite{jl,dyson1}.
  It admits the representation
\bea
\label{sec3.18}
{\tilde{\cal F}}_j(u)=\int ds\int d^4u'{\bf \Sigma}(x,u-u',s)
\eea
The function ${\bf \Sigma}(x,u-u',s)$ vanishes unless the
hyperboloid $(u-u')^2=s$ lies in the region $u^2\ge 0$; otherwise it is
arbitrary. We recall that ${\tilde C}_j(x,\zeta)$ is local in $x$ for
 every fixed $\zeta$. Moreover, from the  Fourier
transform of ${\tilde C}_j(x,\zeta)$, we know that $u\in V^+$ which
is conjugate to $\zeta$.  \\

It is necessary to discuss further the analyticity properties in
the present context. Let us complexify $\zeta$, i.e. define
${\tilde\zeta}=\zeta-i\alpha,~\alpha\in V^+$. We employ the arguments
developed in the previous section to consider the complex $\zeta$-plane.
This defines the forward tube,
$T({\tilde\zeta})$. Thus, ${\tilde C}_j(x,\zeta)$ is boundary
value of an analytic function ${\cal G}_j({x, \tilde\zeta})$ such that
\bea
\label{sec3.19}
{\tilde C}_j(x,\zeta)=
{\rm lim}_{\alpha\rightarrow 0} {\cal G}_j(x,\zeta-i\alpha)
\eea
We obtain the corresponding extended tube $T'(x,{\tilde\zeta})$
by implementing orthochronous complex Lorentz transformations,
$SL_+(2{\bf C})$,  on the 
points of the forward tube $T({\bar\zeta})$. Moreover,
\bea
\label{sec3.20}
 {\cal G}_j(x,{\tilde\zeta})=
0~ {\rm for}~  {\tilde\zeta}^2<0,~{\tilde\zeta}~{\rm real}
\eea
Note that  the real points are separated by spacelike distance.Therefore,
${\tilde C}_j(x,\zeta)$ and ${\tilde C}_j(x, -\zeta)$ are
 analytic continuation of each other.\\
We may interpret this conclusion as a proof of crossing in the following sense. (i) The two local operators
${\tilde C}_j(x,\zeta)~{\rm and}~{\tilde C}_j(x,-\zeta)$  coincide
at the Jost point. (ii) Now we consider the difference of two
Wightman functions $W_4(x_1,x_2,x_3,x_4)- W'_4(x_1,x_3,x_2,x_4)$ which is
\bea
\label{sec3.21}
<0|\phi(x_1)\phi(x_2)\phi(x_3)\phi(x_4)|0> -
<0|\phi(x_1)\phi(x_3)\phi(x_2)\phi(x_4)|0>
\eea
Once we implement OPE for the fields $\phi(x_2)\phi(x_3)$ and
 $\phi(x_3)\phi(x_2)$ we arrive at the JLD representation
and finally see that two functions coincide for  spacelike separated points.
 \\

\bigskip

\noindent{\bf 3.2. PCT Theorem and Conformal Bootstrap Equation}.\\
\bigskip

\noindent We proceed to derive bootstrap equation by invoking PCT
theorem and its equivalence with weak local commutativity as alluded to
in the beginning of this section. Let us briefly recapitulate important
aspects of PCT theorem for motivations.
The PCT theorem is very profound. The theorem,
known as Pauli-L\"uder theorem,   was proved by Pauli \cite{pauli} and
by L\"uder \cite{luder, gr}. 
Furthermore, when the theorem was proved by Pauli and L\"uder, they had  not
considered the possibilities of the violation of discrete symmetries: P, C and
T.  Moreover, in their proof, they considered Lagrangian field theories.
The violation of parity in weak interactions was proposed by Yang and Lee
 \cite{yl} later. The parity violation was experimentally observed very soon
by Wu and collaborators \cite{wu}. Therefore, with hindsight, we may
say that the proof of Pauli-L\"uder PCT theorem  did not
consider the most general case at that juncture. Jost \cite{jost} proved PCT
theorem rigorously for axiomatic local field
theories and established the equivalence of the theorem with weak local
commutativity of Wightman functions.  Dyson \cite{dyson58} as well as Ruelle \cite{ruelle59} 
have further investigated consequences of WLC and analyticity properties
of Wightman functions.
Greenberg \cite{greenberg}
has proved that violation of PCT theorem by Wightman functions
implies violation of Lorentz
invariance of the local field theory.  One of the consequences
of the PCT theorem is that the masses of particle and antiparticle
be equal. The best experimental test is from the $K^0$ and ${\bar K}^0$
mass difference \cite{pdg}:
$-4\times 10^{-19}~GeV<m_{K^0}-m_{{\bar K}^0}<+4\times 10^{-19}~GeV$.
It is obvious why so much of premium is placed
on CPT theorem.
Let us consider the four point Wightman function to
derive the conformal bootstrap equation from the PCT theorem.\\

The  discrete spacetime transformations: parity, P, and time reversal, T,
have following properties. Under P, $(t,{\bf x})\rightarrow (t, -{\bf x})$, 
whereas under time reversal, T, $(t, {\bf x})\rightarrow (-t,{\bf x})$. All
the additive quantum numbers characterizing a field reverse their signs under
C.  Moreover, T is an antilinear operator. The product of the three
discrete operators is denoted as 
${\bf \Theta}=PCT$. For local field theories, if it is invariant under proper
Lorentz transformation, the existence of $\bf {\Theta}$ can be proved. Consider
a complex scalar field, $\Phi(x)$. The action of the operator is
\bea
\label{pct1}
{\bf\Theta}\Phi(x){{\bf\Theta}}^{-1}= \eta{\Phi(-t,-{\bf x})^{\dagger}},~~|\eta|^2=1
\eea
$\eta$, satisfying the constraint, 
is a phase and it is so chosen that the theory is PCT invariant. A Wightman
function of complex scalar field $\Phi(x)$
transforms as follows under $\bf \Theta$
\bea
\label{pct2}
<0|\Phi(x_1)\Phi(x_2)...\Phi(x_n)|0>\rightarrow&&<0|\Phi(-x_1)\Phi(-x_2)...\Phi(-x_n)|0>^*
\nonumber\\&&=<0|\Phi(-x_n)\Phi(-x_{n-1})...\Phi(-x_1)|0> 
\eea
We deal with a real scalar conformal field, $\phi(x)$.
We recall that the four point function,
$W_4(x_1,x_2,x_3,x_4)=<0|\phi(x_1)\phi(x_2)\phi(x_3)\phi(x_4)|0>$
which  transforms as follows
under the PCT operation
\bea
\label{sec3.22}
<0|\phi(x_1)\phi(x_2)\phi(x_3)\phi(x_4)|0>\rightarrow
<0|\phi(-x_4)\phi(-x_3)\phi(-x_2)\phi(-x_1)|0>
\eea
If the theory is invariant under PCT symmetry then
\bea
\label{sec3.23}
<0|\phi(x_1)\phi(x_2)\phi(x_3)\phi(x_4)|0>=
 <0|\phi(-x_4)\phi(-x_3)\phi(-x_2)\phi(-x_1)|0>
\eea
We have noted, in the previous section, that  $W_4(x_1,x_2,x_3,x_4)$
depends on difference of the coordinates,
$y_j=x_j-x_{j+1}$  due to translational invariance.
Recall our definition of  complexified coordinates,
 $\xi_j=y_j-i\eta_j, \eta_j\in V^+$.
 We argue, as before,
that $W_4(\{y_j \}), j=1,2,3$ is the boundary value of
the analytic function ${\cal W}(\{\xi_j \}), j=1,2,3$ of complex variables
$\{\xi_j \}$ and $\xi_j\in T_3$, the forward tube.
We remind that $W_4(\{\xi_j \})$ is  also invariant under orthochronous
complex Lorentz
transformation $SL_+(2{\bf C})$. The set of points  of
 $\xi_j\in T_3$ which are generated under arbitrary  complex Lorentz
transformations, $\Lambda\in SL_+(2{\bf C})$ define an
extended tube $T_3'$. In other words,  the points
 $\{\Lambda\xi_j \}$ obtained from $\{\xi_j \}$ belong to the extended tube.
Moreover, there is a single valued continuation of ${\cal W}_4(\{\xi_j \}$
 to the extended tube \cite{hw}.  We have emphasized
before  that $T_3$ contains only the complex points of the
 forward tube. On the other hand $T_3'$  contains the real points,
$\{y_j \}$, as well. These points are spacelike; the Jost points.
 The Jost points are spacetime points in which
all convex combinations  of successive differences are spacelike.
A Jost point is an ordered set $(x_1,x_2,x_3,x_4)$.\\
{\bf\it The Jost theorem \cite{jost}}:
 {\it A real point of} $\xi_1,\xi_2,\xi_3$
{\it
lies in the
extended tube}, $T_3'$, {\it if and only if all real four vectors of the form}
$\sum_1^{3}\lambda_j\xi_j^{\mu},~\lambda_j\ge 0,~\sum_1^{3}\lambda_j=1$
{\it are
spacelike}  i.e. $(\sum_1^{3}\lambda_j\xi_j^{\mu})^2<0,~\lambda_j\ge 0,~
\sum_1^{3}\lambda_j=1$. \\

\noindent
The equivalence between PCT theorem and WLC: \\
If the PCT theorem holds for
all $x_1,x_2,x_3,x_4$ then for every $x_1,x_2,x_3,x_4$ such that each of the 
$y_j=x_j-x_{j+1}$
is a Jost point. The   WLC condition leading  to 
\bea
\label{sec3.24}
<0|\phi(x_1)\phi(x_2),\phi(x_3)\phi(x_4)|0>=
<0|\phi(x_4)\phi(x_3),\phi(x_2) \phi(x_1)|0>
\eea
is satisfied. The converse statement of   the Jost theorem is paraphrased as
{\it if WLC  holds in a real neighborhood of (\ref{sec3.24}), a
Jost point, then
the PCT condition (\ref{sec3.23}) is valid everywhere}.
Moreover,  WLC implies
validity of PCT symmetry for the conformal scalar.
We are in a position to address the conformal bootstrap proposal in the
present perspective and go through the following steps.
 \\

\noindent 
{\bf I}. The validity of the PCT theorem is {\it assumed} for conformal field
theory. Moreover, we know that
 ${\cal W}_4(\xi_1,\xi_2,\xi_3)$ is a holomorphic function and
(\ref{sec3.24}) holds in the extended tube $T_3'$. Furthermore,   the
four point function is boundary value of an analytic function
\bea
\label{sec3.25}
{\rm lim}_{\eta_j\rightarrow 0}{\cal W}_4(\xi_1,\xi_2,\xi_3)=W_4(y_1,y_2,y_3)
\eea
We also know from \cite{hw}   that 
 ${\cal W}_4(\xi_1,\xi_2,\xi_3)$ is invariant under proper
complex
Lorentz transformations, $SL_+(2{\bf C})$:
$\{\xi_i \}\rightarrow{\Lambda\{\xi_i} \},~ \xi_i\in T_3'$. Let us choose a
$\Lambda$ such
that the set of  complex four vectors $\xi_i^{\mu}\rightarrow -\xi_i^{\mu}$.
Consequently,
\bea
\label{sec3.26}
{\cal W}_4(\xi_1,\xi_2,\xi_3)={\cal W}_4(-\xi_1,-\xi_2,-\xi_3)
\eea
{\bf II}.  We recall that the $r.h.s.$ of the equation
 (\ref{sec3.23}) is a statement of PCT invariance of the four point function.
 It    is also boundary
value of an analytic function.
\bea
\label{sec3.27}
{\rm lim}_{\eta_j\rightarrow 0} {\cal W}_4(\xi_3,\xi_2,\xi_1)=W_4(y_3,y_2,y_1)=
<0|\phi(-x_4)\phi(-x_3)\phi(-x_2)\phi(-x_1)|0>
\eea
{\bf III}. Now consider the difference of two four point functions:
${\cal W}_4(\xi_1,\xi_2,\xi_3) -{\cal W}_4(\xi_3,\xi_2,\xi_1)$.
This is holomorphic in the domain $T_3'$. We know that 
 this difference vanishes  for
${\rm Re}~ \xi_i ,i = 1,2,3 $  by CPT theorem (\ref{sec3.23}).
Let us  invoke  the edge-of-the-wedge theorem \cite{bot,epstein}.
  The essence of the theorem is the following
\cite{epstein,tomozawa}.
Consider two real open cones, $C_1$ and $C_2$, in $R^n$ for generality.
Let the functions $f_1(z)$ and $f_2(z)$, with $z=x+iy$ (note $x,y$ carry
no indices and they have no relationship with $\{y^{\mu}_i \}$ defined above) 
satisfy following properties: (i) The two functions, $f_1(z)$ and $f_2(z)$,
are defined and analytic in the intersection of the tube over 
$C_{\alpha}, \alpha=1,2$; that is $z:~Im~z\in C_{\alpha}$ and also 
analytic in a certain neighborhood of $z=0$. (ii) When $y$ tends to $0$ from 
inside $C_{\alpha}$ the two functions $f_1(x+iy)$ and $f_2(x+iy)$ tend 
to distributions $T_1(x)$ and $T_2(x)$ respectively; in $D_N$ where
$N$ is certain real neighborhood of the point $z=0$. Recall that the
Wightman function, a distribution, is boundary value of the analytic function.
And (iii) $T_1=T_2$. Then $f_1(z)$ and $f_2(z)$ have a common analytic
extension $f(z)$ on the intersection of the neighborhoods of $z=0$ and
the convex closure of $C_1\cup C_2$. Consider a situation where
$C_1$ and $C_2$ are completely opposite i.e. $C_1\cap (-C_2)$ contains an 
open cone, then $f(z)$ is analytic in a neighborhood of $z=0$. It
follows, in our context, that if  the coordinate differences,
$x_i-x_{i+1}, i=1,2,3,4$ are  all spacelike, the two Wightman functions 
are analytic at the real points $\{y_i\}, i=1,2,3$ and they are equal.
Consequently, they are analytic continuations of each other. 
\\

 Therefore, PCT theorem and WLC
 together
with the edge-of-the-wedge theorem will be crucial
to what follows. The immediate conclusion is
\bea
\label{sec3.28}
{\cal W}_4(\xi_1,\xi_2,\xi_3)={\cal W}_4(\xi_3,\xi_2,\xi_1)
\eea
The converse of the above  statement is the following. It is
a consequence of  Hall and Wightman  theorem \cite{hw} that 
 if (\ref{sec3.28}) holds good
in an arbitrary neighborhood of $T_3'$ it also holds good in the extended tube.
Moreover, if it is also valid for
passing into the boundary in the tube $T_3$ then we recover the condition
of PCT
invariance (\ref{sec3.23}). In the historical context, note that the 
first proof of
the edge-of-the-wedge theorem \cite{bot} was presented to prove
dispersion relations for pion-nucleon scattering. The amplitude
was obtained by adopting the LSZ \cite{lsz} reduction technique. Subsequently,
the matrix element of causal commutator of the source currents was
envisaged. Thus the equations of motion were implicitly used. For the present
case we are dealing with Wightman functions and the edge-of-the-wedge theorem
is invoked to prove (\ref{sec3.28}). Thus the  conclusion
is that  PCT invariance is
equivalent to  WLC in CFT.
It follows from equations (\ref{sec3.26}) and   (\ref{sec3.28}) that
\bea
\label{sec3.29}
{\cal W}_4(\xi_1,\xi_2,\xi_3)={\cal W}_4(-\xi_1,-\xi_2,-\xi_3)
\eea
{\it Remark}: Let us try to pass to the boundary in the above equation
for any set of $y_1,y_2,y_3$ in (\ref{sec3.29}). A problem arises. We
shall not be able to obtain a relation between the two functions (in the
above equation) for any set of $\{y_i \}$ for the following reason.
On the $l.h.s.$ $\xi_1,\xi_2,\xi_3$ approach the real  vectors which are
in $V^+$. Note that the real vectors of $-\xi_1,-\xi_2,-\xi_3$ would be
in $V^-$. The equality holds for for ${\rm Re}~(\xi_1,\xi_2,\xi_3)$
and ${\rm Re}~(-\xi_1,-\xi_2,-\xi_3)$ when we are at the Jost point.
\\

Notice the important fact: at the real point of holomorphy,
at the Jost point,  we have the following relation
\bea
\label{sec3.30}
 W_4(\xi_1,\xi_2,\xi_3)=&&<0|\phi(x_1)\phi(x_2)\phi(x_3)\phi(x_4)|0>\nonumber\\
&&
=W(-\xi_3,-\xi_2,-\xi_1)=<0|\phi(x_4)\phi(x_3)\phi(x_2)\phi(x_1)|0>
\eea
This equation has important implications for the conformal bootstrap proposal.
\\

Let us proceed to envisage  the four point Wightman function
$<0|\phi(x_1)\phi(x_2)\phi(x_3)\phi(x_4)|0>$. We  employ the conformal
partial wave expansion. Introduce a complete set of states, $\{|\Psi> \}$,
between the product of the two pair of operators:
$\phi(x_1)\phi(x_2)$ and $\phi(x_3)\phi(x_4)$. The states $\{|\Psi> \}$
belong to the full Hilbert space, ${\cal H}$. Therefore, all irreducible
representations of the conformal group are included as 'intermediate' states.
Now  
\bea
\label{sec3.31}
W_4=<0|\phi(x_1)\phi(x_2)\phi(x_3)\phi(x_4)|0>={\bf\sum}_{|\Psi>}
<0|\phi(x_1)\phi(x_2)|\Psi><\Psi|\phi(x_3)\phi(x_4)|0>
\eea
The sum $\sum |\Psi>$ also includes integration over coordinates as
is the custom when we insert complete set of intermediate states.
We resort to  the ${\rm state}\leftrightarrow{\rm operator}$
correspondence
and then interpret $<0|\phi(x_1)\phi(x_2)|\Psi>$ as a three point function
$<0|\phi(x_1)\phi(x_2){\hat\Psi}|0>$.
Let us  identify $|\Psi>={\hat \Psi}|0>$; ${\hat \Psi}$ represents
the complete set of operator belonging to irreducible representations
of the conformal group. Notice that 
 the second matrix element of the $r.h.s.$ of
the above equation is another three point function and
 $<\Psi|=<0|{\hat{\bar \Psi}}$; ${\hat{\bar\Psi}}$
is the adjoint of ${\hat\Psi}$.
We may reexpress (\ref{sec3.31}) as
\bea
\label{sec3.32}
<0|\phi(x_1)\phi(x_2)\phi(x_3)\phi(x_4)|0>=
{\bf\sum}_{{\hat\Psi},{\hat{\bar\Psi}}}
{\bf\sum}_{\alpha\beta}
\lambda^{\alpha}_{\phi_1\phi_2}
{\cal W}^{\alpha\beta}_{\phi_1\phi_2{\hat\Psi}{\bar{\hat\Psi}}\phi_3\phi_4}
\lambda^{\beta}_{\phi_3\phi_4}
\eea
where $\lambda^{\alpha}_{\phi_1\phi_2}$ and $\lambda^{\beta}_{\phi_3\phi_4}$
 can be read off from the above equation. We have adopted the notation that
$\phi_i$ stands for $\phi(x_i)$ for brevity.
Moreover,
${\cal W}^{\alpha\beta}_{\phi_1\phi_2{\hat\Psi}
{\bar{\hat\Psi}}\phi_3\phi_4}$ is the conformal partial waves (CPW)
\cite{ferrara2,rev1,rev2,rev3}.  The sum over complete set of operators include
all the allowed irreducible representations of the conformal group such
as Lorentz spin, scale dimensions etc.
We have discussed the analyticity and its close relationship with causality
in the previous section in the Wightman's formulation for CFT.
Let us briefly recall the analyticity properties of the commutator of
a pair of composite operators that appear in OPE. 
We 
consider the CPW expansion of another familiar four point function 
\bea
\label{sec3.33}
<0|\phi(x_4)\phi(x_3)\phi(x_2)\phi(x_1)|0>=
{\bf\sum}_{{\hat\Psi},{\bar{\hat\Psi}}}
{\bf\sum}_{\alpha\beta}
\lambda^{\alpha}_{\phi_4\phi_3}
{\cal W}^{\alpha\beta}_{\phi_4\phi_3{\hat\Psi}{\bar{\hat\Psi}}\phi_2\phi_1}
\lambda^{\beta}_{\phi_2\phi_1}
\eea
The two expressions for Wightman functions
(\ref{sec3.32}) and (\ref{sec3.33}) are equal at those Jost points and
these are conformal bootstrap conditions
\cite{ferrara2,polya2, rev1,rev2,rev3}.
\\

We remind that the Wightman functions, satisfying (\ref{sec3.32}) and
(\ref{sec3.33}) are analytic functions in the extended tubes. Let us invoke
Jost's theorem \cite{jost} and Dyson's arguments \cite{dyson58} for the
proof of analyticity of the Wightman functions for real $\{\xi_i \}$ and
when the point corresponds to Jost point. It follows from WLC condition
that
\bea
\label{sec3.34}
{\bf\sum}_{{\hat\Psi},{\hat{\bar\Psi}}}
{\bf\sum}_{\alpha\beta}
\lambda^{\alpha}_{\phi_1\phi_2}
{\cal W}^{\alpha\beta}_{\phi_1\phi_2{\hat\Psi}{\bar{\hat\Psi}}\phi_3\phi_4}
\lambda^{\beta}_{\phi_3\phi_4}
=
{\bf\sum}_{{\hat\Psi},{\bar{\hat\Psi}}}
{\bf\sum}_{\alpha\beta}
\lambda^{\alpha}_{\phi_4\phi_3}
{\cal W}^{\alpha\beta}_{\phi_4\phi_3{\hat\Psi}{\bar{\hat\Psi}}\phi_2\phi_1}
\lambda^{\beta}_{\phi_2\phi_1}
\eea
This is a conformal bootstrap equation. We may discuss an $s-u$ crossing
relation and obtain a bootstrap equation. The point is to note 
that two corresponding Wightman functions are permutation of field
configuration of one another as alluded to in Section 2.3. The
four point function of interest is 
$W_4(x_1,x_2,x_4,x_3)= <0|\phi(x_1)\phi(x_2)\phi(x_4)\phi(x_3)|0>$.
If we follow the known  prescription we obtain the following  equation.
\bea
\label{sec3.35}
<0|\phi(x_1)\phi(x_2)\phi(x_4)\phi(x_3)|0>=
\sum _{{\hat\zeta}}
<0|\phi(x_1)\phi(x_2){\hat\zeta}><{\hat\zeta}|\phi(x_3)\phi(x_4)|0>
 \eea
As before, ${\hat\zeta}$ is a complete set of operators 
created from a complete set of intermediate set of states. 
Notice that at the Jost point,  $(x_3-x_4)^2<0$ the two equations
(\ref{sec3.31}) and (\ref{sec3.35}) are equal. Now the bootstrap equation,
at the Jost point, is
\bea
\label{sec3.36}
{\bf\sum}_{\Psi,{\hat\Psi}}
{\bf\sum}_{\alpha\beta}
\lambda^{\alpha}_{\phi_1\phi_2}
{\cal W}^{\alpha\beta}_{\phi_1\phi_2{\hat\Psi}{\bar{\hat\Psi}}\phi_3\phi_4}
\lambda^{\beta}_{\phi_3\phi_4}=
\sum_{{\zeta}{\hat\zeta}}\sum_{\gamma\delta}\lambda^{\gamma}_{\phi_1\phi_2}
{\cal W}^{\gamma\delta}_{\phi_1\phi_2{\zeta}{\hat\zeta}\phi_4\phi_3}
\lambda^{\delta}_{\phi_4\phi_3}
\eea
Each expression on $l.h.s$ and $r.h.s$ of the above
equation corresponds to a boundary values of an analytic function.
Moreover, each analytic function is defined in a domain corresponding to 
its extended tube. The two analytic functions coincide at the Jost point
as expressed in (\ref{sec3.36}). Thus by invoking the edge-of-the-wedge
theorem we conclude that the two (Wightman) functions are analytic
continuation of each other. 
\\

 We have alluded to the
importance and significance  of this equation. Let us deliberate
on a few points. Notice that the equation holds when points are separated
by spacelike distance. If we considered Fourier transform of the two
Wightman functions, $W_4(\xi_1,\xi_2,\xi_3)$ and $W_4(\xi_3,\xi_2,\xi_1)$
the conjugate momenta are spacelike i.e. for $\xi_3^2<0$. 
Therefore, the two functions
coincide in the region where momenta are spacelike. This is the situation
in case of the absorptive amplitudes of $s$ and $u$ channels of
a scattering amplitude derived from the LSZ reductions. The  absorptive
parts of the two amplitude are equal when momenta are real and lie in
the unphysical region. It was necessary to prove that the two absorptive parts
are analytic continuation of each other. The situation is
 the same here. It is
necessary to identify the domain of analyticity of each of the
Wightman functions and argue that they are analytic continuation of each
other. We have proved this through the chain of arguments presented earlier.
We have persuasively argued in the previous section that a pair of Wightman
functions are holomorphic in the union of their extended tubes.
We think that the PCT theorem together with its equivalence of WLC
provide a very strong basis to derive the conformal bootstrap equation.
It is needless to emphasize the crucial role played by  the
edge-of-the-wedge theorem.\\

\noindent
{\it Remark:} The  above bootstrap condition is not specifically valid
for a nonderivative scalar conformal
field theory. If we consider four point Wightman function, for
nonderivative conformal fields,  which belong to finite dimensional irreducible
representation of conformal group then the above proof will go through
with appropriate modifications. Now the
corresponding Wightman function will carry tensor indices as the fields
would transform according to the representations of
$SL(2{\bf C})\otimes SL(2{\bf C})$ and they will carry their
conformal dimensions \cite{fp2}.
 Thus the $W_4$ will transform covariantly
under $SL(2{\bf C})\otimes SL(2{\bf C})$. The preceding arguments will
essentially go
through with adequate technical modifications only.
 Consequently, the analyticity properties and bootstrap equations
will continue to hold for the general four point functions as long as the
fields are of nonderivative types.
Therefore, we conclude that the two resulting  four point functions,
in general,
will be analytic continuation of each other.
\\

To briefly summarize the contents of this section: (i) We considered OPE
of a pair of nonderivative scalar conformal fields and expanded them in
a set of composite (also nonderivative) conformal fields. Next, we considered
OPE for the commutator of the pair of fields and noted that the commutator
vanishes when they are separated by spacelike distance. Then we argue
that the matrix element of the Fourier transforms of the difference of
the two composite field operators enjoy certain support properties.
Consequently,  a representation, for the matrix elements could derived
which is analogous to Jost-Lehmann-Dyson representation. The analyticity
properties of the matrix elements are discussed. The connections with
bootstrap equation are presented.\\

In the second part, we invoked PCT theorem to aim at derivation of the
bootstrap equation. The equivalence of weak local commutativity (WLC)
with the PCT theorem plays a crucial role in arriving at the bootstrap
equation. We went through a number of steps. It is important, in our view,
to note that although the two four point Wightman functions coincide
at the Jost point, one has to demonstrate
 that the two four point functions are
analytic functions. It is necessary to identify their domain of holomorphy
and then invoke the edge-of-the-wedge-theorem to show that these two
functions are analytic continuation of each other. \\

\bigskip

\noindent
{\bf {4. Summary and Conclusions }.}\\

\bigskip

The objective of this investigation was to  study  analyticity and
 crossing properties of four point correlation functions of
conformal field theory. We  adopted Wightman's formulation.
It is well known that Wightman axioms are not respected by all conformal
field theories. Therefore, we considered 
a nonderivative real scalar field, $\phi(x)$,
 which has the desired properties \cite{mack1}. We invoked the arguments 
that Wightman functions are boundary values of analytic functions of
several complex variables and  focused on the four point
 function. The primitive domain of analyticity of the corresponding analytic
 function was identified to be the forward tube, $T_3$ since the four point
function depends on three independent coordinates due to translational
invariance.    Next, the extended tube,
 $T_3' $ was defined following the standard procedure where the 
function is analytic and is single valued.
  A permuted four point Wightman function
was considered.
The two four point functions  are related by crossing.
As a prelude, we presented analyticity and crossing properties of the
three point function in some detail incorporating essential results of
\cite{mpla}.\\

 Moreover, we have studied the analyticity properties
of three point function in the momentum space representation. The work
of Jost \cite{jost31} was crucial.  The 
 R-product of three point function was shown to be  related to 
Wightman functions. We utilized this fact to investigate analyticity of
momentum space three point function.\\

Next, we analyzed the crossing properties of four point function. 
In general, each of the permuted Wightman function is boundary value of
the corresponding analytic function. The domain
of analyticity is union of the domain of holomorphy of  four point
functions. However,  the entire  domain of holomorphy of the
collection all the Wightman function might be much larger. 
We have considered a pair of Wightman functions
 at a time and then found the domain of holomorphicity.
The Fourier transforms of the two four point functions were considered and
 we  read off their support properties in the momentum
 space.  We adopted the prescriptions of
\cite{mpla} to derive representation of the Fourier transform
of the four point Wightman function. This representation is similar to
the Jost-Lehmann-Dyson representation derived for the matrix element
of causal current commutators in field theory. 
It might be useful to recall a few important results on analyticity
of four point scattering amplitude from axiomatic field theory view point.
One of the most important result is the existence of the Jost-Lehmann-Dyson
representation \cite{jl,dyson1}. 
This is derived from microcausality and temperedness of
the amplitude. The next important result for the proof of the existence
of Lehmann ellipses which crucially depends of Jost-Lehmann-Dyson 
result. Moreover, the polynomial boundedness of the nonforward scattering
amplitude follows from the linear program without invoking unitarity of
S-matrix. It is, sometimes, stated that polynomial boundedness and
existence of Lehmann ellipses are underlying assumptions in the proof of
dispersion relations. However, it is not the case - the two properties are 
proved from general axioms.
 Martin invoked the nonlinear constraint (unitarity
of S-matrix) to demonstrate the existence of enlarged domain of analyticity,
i.e. existence of an ellipse known as Lehmann-Martin ellipse. The celebrated
Froissart bound is derived rigorously from axioms of general field theory
without any extra assumptions. In other words, polynomial boundedness of
the amplitude, existence of Lehmann ellipses and enlarged domain of analyticity
(the existence of Martin ellepse) have been proved. Consequently, the 
fixed-t dispersion relations are proved in the domain mentioned above. The
purpose of this digression is to discuss analyticity of the Wightman function
(say four point function) in the light of preceding remarks. First of
all, in conformal field theories,  we deal with Green functions and not
scattering amplitudes. The 'external' legs are off the mass shell. Polykov
\cite{polya2} has discussed this aspect very succintly while deriving
his bootstrap equation. All the external legs have large spacelike momenta.
The momentum transfers are spacelike and he assigns large spacelike
values to  some of them while computing discontinuity to obtain the desired
equation. In order to derive a dispersion relation, we propose the 
following strategy. Bogoliubov \cite{book6} had proved fixed-t dispersion
relation by taking external legs to large spacelike region and then 
analytically continued the function to onshell values. It might be worth while
to adopt his prescription. However, we are aware that we do not deal with a 
scattering amplitude, there is no passage to the mass shell in a rigorous
manner. We may conjecture, optimistically, that a dispersion relation might 
be proved for the conformal field theory in future. We turn to crossing
symmetry.  
Indeed, starting
from this point, we get insights into crossing symmetry for a pair
of permuted Wightman function at a point where the two functions
coincide.  Next
step was to derive the domain of analyticity in the coordinate space.
We depended on two important results.  First, the Hall-Wightman theorem
was invoked to argue that the analytic function depend on Lorentz
invariants constructed from the complexified four vectors. The second
ingredient was to appeal to the Jost theorem. As far as we are aware,
the envelope of holomorphy have not been constructed for four point
Wightman functions completely in QFT \cite{kt,kallen,russian} i.e. as
exhaustively as for the three point function \cite{kw}.
  Our principal goal was to establish crossing
for four point functions. Therefore, it suffices, for our purpose, to
consider a pair of Wightman functions and identify the domain of
holomorphy.  Consequently, we can proceed to prove 
crossing for Wightman functions taken pairwise. There was one more
technical obstacle. In case of the three point functions there were only
two Hall-Wightman Lorentz invariant complex variables for all 
practical purposes
(although there are three of them $z_{11},z_{22}, z_{12}$; and crossing was
proved \cite{mpla}).  Note that
for the four point functions, the number of Hall-Wightman variables
increase and  the prescription of \cite{mpla}  does
not go through in a straight forward manner.
 Therefore, we simplified  the task a little bit and
derive constraints on (real) Hall-Wightman invariants since we go
to a domain where Jost theorem is applicable. Thus in Section 2,
we  established crossing for a pair of permuted Wightman functions.\\

The third section was devoted to study analyticity properties of
the matrix elements of composite operators which arise in the OPE.
In this section the microcausality plays a crucial role in establishing
the analyticity properties. The OPE of a pair of nonderivative
conformal fields has been investigated by Mack \cite{mack1}.
In the  OPE, the composite fields are also of nonderivative type and they 
belong to the irreducible representations of the conformal group. Consequently,
they respect Wightman axioms. We considered, OPE of the commutator of the
two scalar fields. Thus we obtain difference of two composite
fields  as has ben demonstrated in section 3.
We considered the Fourier transforms  and then took matrix elements
of the composite  operators. This matrix element is constrained from
the microcausality arguments. The constraints are the same as those
utilized to derive the Jost-Lehmann-Dyson representation (JLD) \cite{jl,dyson1}. 
We recall that,
 JLD representation was crucial to derive crossing in QFT. Moreover, we use
the techniques developed in section 2 to study the analyticity
properties of these matrix elements. 
We presented a  derivation of the conformal  bootstrap equation
from a novel perspective. We invoked two very powerful
theorems of axiomatic field theory to accomplish our goal. 
 We first appeal to  the PCT theorem. PCT theorem is profound
and is respected by all local field theories which obey Wightman axioms.
We noted that  two Wightman functions are equal if they are PCT transform
of each other.
We utilized  the equivalence of
PCT theorem and weak local commutativity which was rigorously proved by Jost.
 Therefore, we were able to relate two Wightman functions
at Jost point. Next, we presented a  series of steps to relate the 
two Wightman functions. The conformal partial wave expansion
technique was applied to arrive at the conformal bootstrap equation.
It is not adequately emphasized that the equality between two four point
functions holds when a pair of coordinates, for their real vales,
 are separated by spacelike distance. It is essential to prove that
the four point functions are analytic at that point. Moreover, from the
perspectives of Wightman axioms, it is required that we identify their
respective extended 
domain of holomorphicity of the two functions in the complex domain.
 Dyson \cite{dyson58} and Ruelle \cite{ruelle59}  have  
proved the analyticity of the
Wightman functions at Jost point in the context of WLC \cite{jost}.
 Furthermore, 
as has been argued in section 3, the bootstrap equation is not special
to the case of Wightman function of four scalar field. Indeed,
a four point functions can be defined as product of  four conformal fields
belonging to irreducible representations of the conformal group, in a general
scenario, as long as they satisfy Wightman axioms. Then the corresponding
bootstrap equation holds. It is important to note that three fundamental
theorems of axiomatic local field theories such as
PCT theorem, the theorem stating equivalence between PCT theorem
and weak local commutativity and the edge-of-the-wedge theorem, are
invoked to derive the conformal bootstrap equation rigorously. We presented
arguments for the $s-u$ bootstrap equation. \\

We conclude the article with following remarks.  It is tempting to
suggest that crossing and analyticity of n-point function can be derived
following the techniques introduced here. It must be noted
that proof of crossing and analyticity for n-point functions in 
axiomatic  QFT is a formidable task.
Bros \cite{bros} has comprehensively reviewed the progress on
crossing and analyticity properties of n-point amplitude in axiomatic QFT
at that juncture. Our understanding is that several issues have remained
unresolved in this topic. 
Another avenue to explore, in order to derive crossing relations
for n-point functions in the context of CFT,
 is to follow a clue from QFT. We should look for
a generalization of Jost-Lehmann-Dyson representation for the 
n-point function in CFT. The JLD representation was the principal
ingredient since the coincident region was identified through
this technique. Thus the singularity free region was identified. There
have been attempts, in the past, to obtain integral representations
for the VEV of product of field operators and VEV of the commutators
of string of field operators in the axiomatic QFT 
\cite{streaterjmp,streater1}.
Therefore, those results might be utilized to prove crossing for,
at least, a pair of permuted Wightman functions. However, we have
not established existence of JLD representation to n-point functions in CFT. 
Therefore, it might be premature to speculate
that problem of crossing and analyticity could be solved in a straight
forward manner in CFT for n-point functions.
 However, conformal symmetry is very powerful. 
It is well known that two point and three point functions get fixed
in CFT up to constant coefficients. Moreover, conformal symmetry
imposes constraints on the structure of the n-point functions ($n>3$).
In principle, given the two point and three point functions of a theory,
the structure of n-point functions could be inferred. There are
reasons to be optimistic that CFT might provide more insights into
the analyticity and crossing properties of n-point 
functions.\footnote{ Witten has suggested that it might be worth while to 
investigate questions about analytic continuations
and crossing properties of n-point correlation functions of conformal 
field theories.
    \cite{W}}.\\

\noindent Acknowledgments:  I am thankful to Edward Witten
for very valuable discussions on analyticity properties of scattering 
amplitudes. I was influenced by his suggestions that the correlation functions
of CFT  have  better
prospects of unraveling the analyticity properties. 
The warm hospitality of Professor Witten at the Institute for Advanced Study,
in 2017, is gratefully acknowledged. 
I would like to thank Hadi Godazgar for several
useful remarks and for valuable suggestions to improve the manuscript.

\end{document}